\documentclass{article}

\usepackage{arxiv}

\usepackage[utf8]{inputenc} 
\usepackage[T1]{fontenc}   
\usepackage{hyperref}    
\usepackage{url}           
\usepackage{booktabs}      
\usepackage{amsfonts}       
\usepackage{nicefrac}    
\usepackage{microtype}      
\usepackage{lipsum}
\usepackage{changepage}
\usepackage{graphicx}
\usepackage{amsmath}
\usepackage{amsfonts}
\usepackage{amssymb}
\usepackage{makeidx}
\usepackage{algorithm,algorithmic}
\usepackage[switch]{lineno}
\usepackage{xcolor}
\usepackage{bm}
\usepackage[justification=centering]{caption}
\usepackage{amsfonts}
\usepackage[bbgreekl]{mathbbol}
\DeclareSymbolFontAlphabet{\mathbbm}{bbold}
\DeclareSymbolFontAlphabet{\mathbb}{AMSb}%
\usepackage{amsthm}

\newtheorem{problem}{Problem}
\newtheorem{definition}{Definition}

\newtheorem{lemma}{Lemma}

\usepackage[title]{appendix}%

\title{Data-driven Model Predictive Control using MATLAB}

\author{
  Midhun T. Augustine\\
  Automation Lab\\
  Indian Institute of Technology Bombay, India \\
  midhunta30@gmail.com \\
 \vspace{.01cm}\\
  Date of initial version: 10 - 05 - 2025\\ 
   Date of current version: 10 - 10 - 2025\\ \vspace{.01cm}}

\begin{document}
\maketitle

\begin{abstract}
This paper presents a comprehensive overview of data-driven model predictive control, highlighting state-of-the-art methodologies and their numerical implementation. The discussion begins with a brief review of conventional model predictive control (MPC), which discusses both linear MPC (LMPC) and nonlinear MPC (NMPC). This is followed by a section on data-driven LMPC, outlining fundamental concepts and the implementation of various approaches, including subspace predictive control and prediction error methods. Subsequently, the focus shifts to data-driven NMPC, emphasizing approaches based on neural network models. The paper concludes with a review of recent advancements in data-driven MPC and explores potential directions for future research.
\end{abstract}


\section{Introduction}

Mathematical models are commonly used to represent a particular  problem/system of interest through mathematical equations. Traditionally, such models have been developed using first-principles approaches or domain knowledge, incorporating fundamental laws such as mass and energy balances, heat transfer relations, and related physical phenomena. These models can generally be classified into two categories:
\begin{enumerate}
     \item \textbf{Static model}: describes steady-state or time-independent relationships among variables in the system.
     \item \textbf{Dynamic model}: describes transient or time-dependent relationships among system variables.  Dynamic models can be represented in  continuous-time as differential equations or in discrete-time as  difference equations.
 \end{enumerate}

\textbf{Model predictive control} (MPC) is a modern control approach which is an optimization-based feedback control strategy. The conventional MPC relies on a dynamic model of the system, typically in discrete-time, to predict future behavior (either output or states) over a finite control input sequence \cite{bFA,bLJ}. The predicted behavior is characterized through a cost function, which depends on both the predicted output/state sequence and the control sequence. The control input sequence is then determined by minimizing the predicted behavior (cost function), and the first element of the control sequence is applied to the system. This results in a receding horizon and closed-loop control scheme, where the current input is a function of the current output or state. In conventional MPC, the dynamic model is assumed to be known. However, in many practical applications, the system model may not be available. With recent developments in the fields of data science and machine learning (ML), new approaches have emerged for mathematical modeling and control, emphasizing data-driven models \cite{bLCK20}. This has led to the emergence of data-driven control, a subfield focused on designing control strategies without relying on first-principle models. \textbf{System identification} (SysID), which deals with deriving dynamic models from data, plays a significant role in this domain \cite{bLL99,bGP19}. \textbf{Data-driven MPC} (D-MPC), which integrates SysID, data science, and MPC, is a rapidly growing area in control systems research \cite{bJC19,bJB21,bPV23}. 
\par
\begin{figure} 
 		\begin{center}
 		\includegraphics [scale=.5] {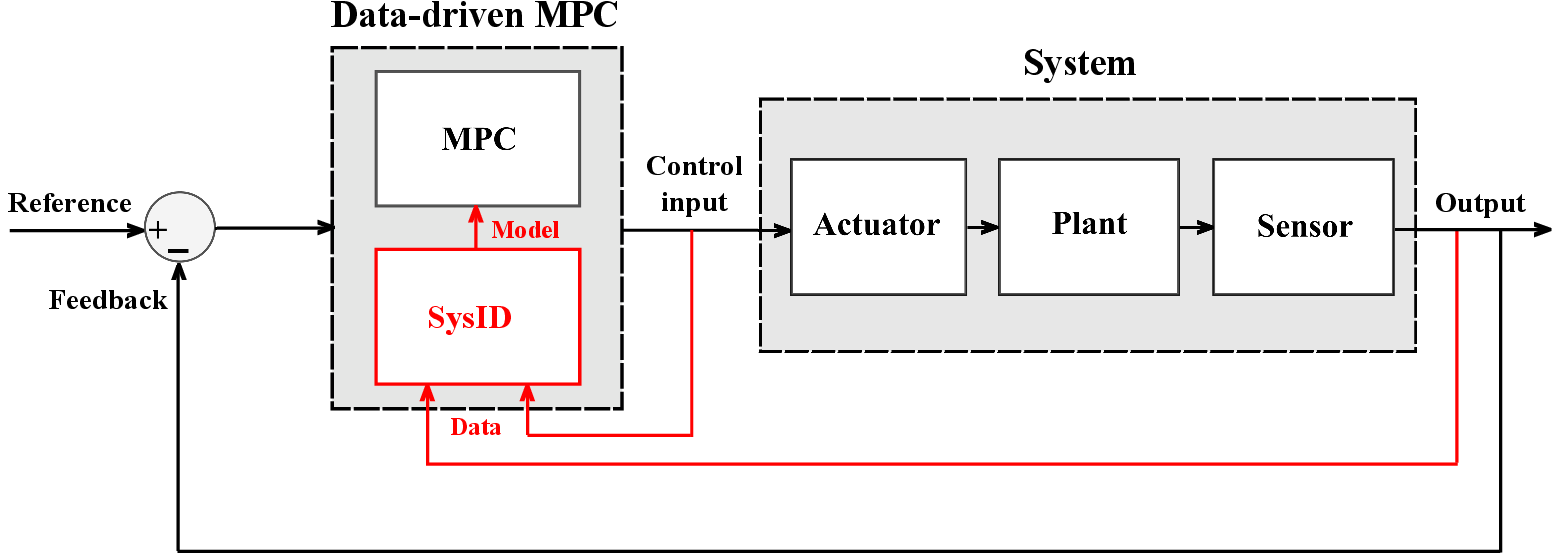}
 		\caption{{  Data-driven MPC block diagram.}}
        \label{figD}
 	\end{center}
 \end{figure}

 D-MPC offers several advantages over conventional MPC schemes. It utilizes available experimental or industrial data, eliminates the need for a first-principle model, and provides adaptive and robust control solutions that are computationally efficient. The general block diagram of a D-MPC implementation is given in Fig. \ref{figD}.   In a typical D-MPC implementation, dynamic models identified from input-output data using SysID methods are employed to predict the system response over a finite prediction horizon.  These approaches, where model identification is the key step, are called model-based D-MPC. Recently, model-free D-MPC schemes have also been introduced in which the data is directly provided to the MPC, bypassing the SysID step (in Fig. \ref{figD}), and the MPC computes the control input \cite{bJB21,bPV23}. As these approaches directly compute the control input from data and do not rely on explicit system models, they are often referred to as direct D-MPC. Based on the nature of the model used for prediction,  D-MPC schemes can be grouped into the following classes:
\begin{enumerate}
    \item \textbf{Data-driven Linear MPC} (D-LMPC):  deals with identifying a linear model from data, which is then used for designing an LMPC scheme. The optimization problem for D-LMPC is normally a quadratic programming problem and will be convex.

    \item \textbf{Data-driven Nonlinear MPC} (D-NMPC):  deals with identifying a nonlinear model from data, which is then used for designing an NMPC scheme. The optimization problem for D-NMPC is normally a nonlinear programming problem and will be non-convex.
\end{enumerate} 
\par \textit{Notations and definitions:} The set of real numbers are denoted by $\mathbb{R}$, the $n$ - dimensional Euclidean space by $\mathbb{R}^{n}$, and 
the space of $m \times n$ real-valued matrices by $\mathbb{R}^{m \times n}$. The identity matrix of order $n\times n$ is denoted by $\mathbf{I}_{n}$. For a matrix $\mathbf{M}\in \mathbb{R}^{m \times n},$ the notation $\mathbf{M}_{(i,:)}$ denotes the $i^{th}$ row, $\mathbf{M}_{(:,j)}$ denotes the $j^{th}$ column, and  $\mathbf{M}_{(i:p,j:q)}$ denotes the submatrix containing rows from $i$ to $p$ and columns from $j$ to $q$. 
The
 $p-$norm of a vector $\mathbf{x}\in \mathbb{R}^{n}$  is defined as ${{  {\parallel \mathbf{x} \parallel}_{p}=({\sum_{i=1}^{n} |x_{i}|^{p}} )^{\frac{1}{p}} }}$. The Frobenious norm of a  matrix $\mathbf{A}\in \mathbb{R}^{m \times n}$ is defined as
${{ {\parallel \mathbf{M} \parallel}_{F}=\sqrt{\sum_{i=1}^{m}\sum_{j=1}^{n}M_{ij}^{2}} = \sqrt{\operatorname{trace}(\mathbf{M}^{\top}\mathbf{M})}. }} $ 
The block Hankel matrix consists of $\mathrm{N} \times \mathrm{H}$ samples of the vector $\mathbf{v}$, starting from the index $k$ is defined as:
\begin{equation}
    \label{eqhv} \mathbf{H}_{(\mathrm{N},\mathrm{H},\mathbf{v}_k)}= {{ \left[\begin{matrix} \mathbf{v}_{k} &  \mathbf{v}_{k+1} & \dots & \mathbf{v}_{k+\mathrm{H}-1} \\ 
    \mathbf{v}_{k+1} &  \mathbf{v}_{k+2} & \dots & \mathbf{v}_{k+\mathrm{H}}\\
    \vdots & \vdots &  & \vdots\\
    \mathbf{v}_{k+\mathrm{N}-1} &  \mathbf{v}_{k+\mathrm{N}} & \dots & \mathbf{v}_{k+\mathrm{N}+\mathrm{H}-2}
\end{matrix}\right] }} =\left[\begin{matrix} \mathbf{V}_{k} & \mathbf{V}_{k+1} & \dots & \mathbf{V}_{k+\mathrm{H}-1}
\end{matrix}\right].
\end{equation}
A general optimization problem can be represented as:
\begin{equation}
\label{eqfopt}
\begin{aligned}
 \underset{\mathbf{z}}{\inf}   ~&~ J(\mathbf{z})\\
 \mathrm{subject \hspace{0.1cm}to} ~&~ \mathbf{z} \in \mathbb{Z}\subseteq \mathbb{R}^{h}
     \end{aligned}
\end{equation}
where $\mathbf{z}$ is the decision vector, $J$ is the cost function (loss function), and $\mathbb{Z}$ is the constraint set. The optimum value of the decision vector is denoted by $\mathbf{z}^{*}$.  
Constraint sets can be defined using linear inequalities:
    $\mathbb{Z}=\{\mathbf{z}\in \mathbb{R}^{n}: \mathbf{F}\mathbf{z}\leq \mathbf{g} \}$ 
    or $\mathbb{Z}=\{\mathbf{z}\in \mathbb{R}^{n}: \mathbf{z} _{min} \leq \mathbf{z} \leq \mathbf{z}_{max} \}$.   The two forms of optimization problems used in this paper are:
    \begin{enumerate}
        \item \textbf{Quadratic programming} (QP) is of the form: 
\begin{equation}
\label{eqqp}
\begin{aligned}
 \underset{\mathbf{z}}{\inf}    ~&~\mathbf{z}^{\top}\mathbf{H}\mathbf{z}+ \mathbf{z}^{\top}\mathbf{q}\\
\mathrm{subject \hspace{0.1cm}to} ~&~ \mathbf{F}\mathbf{z} \leq \mathbf{g}\\
  ~&~ \mathbf{F}_{eq}\mathbf{z} = \mathbf{g}_{eq}.
     \end{aligned}
\end{equation}
\item \textbf{Nonlinear programming} (NLP) is of the form: 
\begin{equation}
\label{eqnlp}
\begin{aligned}
 \underset{\mathbf{z}}{\inf}  ~&~  J(\mathbf{z}) \\
\mathrm{subject \hspace{0.1cm}to} ~&~ \mathbf{f}_{in}(\mathbf{z}) \leq \mathbf{0}\\
  ~&~ \mathbf{f}_{eq}(\mathbf{z}) = \mathbf{0}
     \end{aligned}
\end{equation}
where $J(\mathbf{z})$ is a nonlinear function of the decision vector $\mathbf{z}$, and $\mathbf{f}_{in}(\mathbf{z}),\mathbf{f}_{eq}(\mathbf{z})$ are nonlinear functions used to represent the inequality and equality constraints, respectively.
\end{enumerate}
\par A list of the most prominently used terminologies and symbols in this paper is given below:
\begin{enumerate}
        \item \textbf{Data}: consists of measurements or samples of the variables of interest in the process or problem. In this paper, data refers to the input and output samples used to train the model.

\item \textbf{Model}: is a mathematical equation that relates the variables of interest in the system. In this paper, we are interested in dynamic models that represent the temporal relationships between the variables of interest. 

\item \textbf{State-space model}: is one particular class of dynamic models. State space models are input-state-output models where states act as intermediate variables that capture the internal dynamics of the system. 

\item \textbf{SysID}: deals with the identification of dynamic models from data. This paper focuses on the identification of state-space models using input-output data, which are used to design output or state-based MPC schemes.

\item \textbf{Controller}: deals with controlling the output (dependent variables) by properly adjusting the input (independent variables) to improve the system performance. The control approach we focus on in this paper is MPC. 
    
    \item \textbf{x} -  state vector and $\textbf{x}_{k}$ denotes state at time instant $k$.
    \item \textbf{u} -  control (manipulated) input vector.  
    \item \textbf{y} -  output (controlled) vector.
   \item $L_{y}$ - loss function in model training, which is a function of output prediction error.

   \item $J_{y}$ and $J_{x}$ - cost function for output and state-based MPC.      
\end{enumerate}
In the rest of the paper, we will be discussing major approaches in D-LMPC and D-NMPC, and their implementation with numerical examples. The Matlab codes for the numerical illustration can be accessed via the link \footnote{https://github.com/MIDHUNTA30/D-MPC-MATLAB}. Before moving on to D-MPC, we review conventional LMPC and NMPC in the next section.

 \section{Conventional MPC}
This section begins with a brief review of conventional LMPC that assumes the availability of a linear model to make predictions. Both output and state-based LMPC schemes are explained.  This is followed by a discussion on output and state-based NMPC algorithms, where the key distinction lies in the use of a nonlinear model for predictions.
\subsection{Conventional LMPC}
\label{seclmpc}
Consider a \textbf{linear time-invariant} (LTI) system in discrete-time:
\begin{equation}
\label{eqlin}
\begin{aligned}
\mathbf{x}_{k+1}&=\mathbf{A}\mathbf{x}_{k}+\mathbf{B}\mathbf{u}_{k}\\
\mathbf{y}_{k}&=\mathbf{C}\mathbf{x}_{k}
\end{aligned}
\end{equation}
where $k\in \mathbb{T}$ is the discrete time instant, $\mathbf{y}_{k}\in \mathbb{Y} \subseteq \mathbb{R}^{p}$ is the output (controlled) vector, $\mathbf{u}_{k}\in \mathbb{U} \subseteq \mathbb{R}^{m}$ is the input (control or manipulated) vector,
$\mathbf{x}_{k}\in \mathbb{X} \subseteq \mathbb{R}^{n}$ is the state vector, $\mathbf{A} \in \mathbb{R}^{n \times n}$ is the system matrix, 
$\mathbf{B} \in \mathbb{R}^{n \times m}$ is the input matrix, and
$\mathbf{C} \in \mathbb{R}^{p \times n}$ is the output matrix. The sets $\mathbb{Y},$ $\mathbb{U}$, and $\mathbb{X}$   are the constraint sets for the outputs, inputs, and states, which can be defined as \cite{bMA1}:
\begin{equation}
\label{eqconstr}
\begin{aligned}
   &\mathbb{Y}=\{\mathbf{y} \in \mathbb{R}^{p}:  \mathbf{F}_{\mathbf{y}}\mathbf{y} \leq \mathbf{g}_{\mathbf{y}} \},  \hspace{1cm} \mathbf{F}_{\mathbf{y}} \in \mathbb{R}^{c_{y} \times p}, \mathbf{g}_{\mathbf{y}} \in \mathbb{R}^{c_{y} }\\
    &\mathbb{U}=\{\mathbf{u} \in \mathbb{R}^{m}:  \mathbf{F}_{\mathbf{u}}\mathbf{u} \leq \mathbf{g}_{\mathbf{u}} \},  \hspace{1cm} \mathbf{F}_{\mathbf{u}} \in \mathbb{R}^{c_{u} \times m}, \mathbf{g}_{\mathbf{u}} \in \mathbb{R}^{c_{u}}\\
  & \mathbb{X}=\{\mathbf{x} \in \mathbb{R}^{n}:  \mathbf{F}_{\mathbf{x}}\mathbf{x} \leq \mathbf{g}_{\mathbf{x}} \},  \hspace{1cm} \mathbf{F}_{\mathbf{x}} \in \mathbb{R}^{c_{x} \times n}, \mathbf{g}_{\mathbf{x}} \in \mathbb{R}^{c_{x}}.
  \end{aligned}
\end{equation}
For the linear system in Eq. (\ref{eqlin}), the controllability and observability matrices are defined as:
\begin{equation}
\mathbf{C}_{n}=\left[\begin{matrix}\mathbf{B}&\mathbf{AB}&\dots&{\mathbf{A}}^{n-1}\mathbf{B}\end{matrix}\right] \hspace{1cm}\mathbf{O}_{n}=  {{\left[\begin{matrix}\mathbf{C}\\\mathbf{CA}\\\vdots\\\mathbf{C}{\mathbf{A}}^{n-1}
\end{matrix}\right] }}
\end{equation}
A linear system in Eq. (\ref{eqlin}) is said to be controllable if the controllability matrix is of full row rank, i.e., rank ($\mathbf{C}_{n}$) $= n$. Similarly, the system is said to be observable if rank ($\mathbf{O}_{n}$) $= n$. In linear MPC design, we normally assume that the system for which the MPC control law is designed is both controllable and observable. 
\par Using the state equation in  Eq. (\ref{eqlin}),
the predicted state and output sequences can be computed as:
\begin{align}
 \label{eqxpred}
 &\underbrace{\left[\begin{matrix}
\mathbf{x}_{k+1|k}\\\mathbf{x}_{k+2|k}\\\vdots\\ \mathbf{x}_{k+\mathrm{N}|k}
\end{matrix}\right]}_{\mathbf{X}_{k+1}}=    \underbrace{\left[\begin{matrix}
\mathbf{A}\\ \mathbf{A}^{2}\\ \vdots \\ \mathbf{A}^{\mathrm{N}} 
\end{matrix}\right]}_{\mathbf{A}_{\mathbf{X}}} \mathbf{x}_{k|k}+ \underbrace{\left[\begin{matrix}
\mathbf{B} & \mathbf{0}  & \dots &\mathbf{0}  \\ \mathbf{A}\mathbf{B}& \mathbf{B}  & \dots &\mathbf{0}\\ \vdots & \vdots& & \vdots\\ \mathbf{A}^{\mathrm{N}-1}\mathbf{B} & \mathbf{A}^{\mathrm{N}-2}\mathbf{B}  & \dots & \mathbf{B}   
\end{matrix}\right]}_{\mathbf{B}_{\mathbf{X}}}\underbrace{\left[\begin{matrix}
\mathbf{u}_{k|k}\\\mathbf{u}_{k+1|k}\\\vdots\\ \mathbf{u}_{k+\mathrm{N}-1|k}
\end{matrix}\right]}_{\mathbf{U}_{k}}\\
\label{eqypred}
 &\underbrace{\left[\begin{matrix}
\mathbf{y}_{k+1|k}\\\mathbf{y}_{k+2|k}\\\vdots\\ \mathbf{y}_{k+\mathrm{N}|k}
\end{matrix}\right]}_{\mathbf{Y}_{k+1}}=    \underbrace{\left[\begin{matrix}
\mathbf{C}\mathbf{A}\\\mathbf{C}\mathbf{A}^{2}\\ \vdots \\ \mathbf{C}\mathbf{A}^{\mathrm{N}} 
\end{matrix}\right]}_{\mathbf{A}_{\mathbf{Y}}} \mathbf{x}_{k|k}+ \underbrace{\left[\begin{matrix}
\mathbf{C}\mathbf{B} & \mathbf{0}  & \dots &\mathbf{0}  \\\mathbf{C}\mathbf{A}\mathbf{B}& \mathbf{C}\mathbf{B}  & \dots &\mathbf{0}\\ \vdots & \vdots& & \vdots\\ \mathbf{C}\mathbf{A}^{\mathrm{N}-1}\mathbf{B} & \mathbf{C}\mathbf{A}^{\mathrm{N}-2}\mathbf{B}  & \dots &\mathbf{C}\mathbf{B}   
\end{matrix}\right]}_{\mathbf{B}_{\mathbf{Y}}}\underbrace{\left[\begin{matrix}
\mathbf{u}_{k|k}\\\mathbf{u}_{k+1|k}\\\vdots\\ \mathbf{u}_{k+\mathrm{N}-1|k}
\end{matrix}\right]}_{\mathbf{U}_{k}}=\mathbf{C}\mathbf{X}_{k+1}
\end{align}
where $\mathbf{X}_{k+1} \in \mathbb{R}^{n\mathrm{N}}$, $\mathbf{Y}_{k+1} \in \mathbb{R}^{p\mathrm{N}}$, $\mathbf{U}_{k} \in \mathbb{R}^{m\mathrm{N}}$, $\mathbf{A}_{\mathbf{X}} \in \mathbb{R}^{n\mathrm{N}\times n}$ $\mathbf{B}_{\mathbf{X}} \in \mathbb{R}^{n\mathrm{N}\times m\mathrm{N}}$, $\mathbf{A}_{\mathbf{Y}} \in \mathbb{R}^{p\mathrm{N}\times n}$, $\mathbf{B}_{\mathbf{Y}} \in \mathbb{R}^{p\mathrm{N}\times m\mathrm{N}}$, and $\mathbf{x}_{i|k}, \mathbf{y}_{i|k}, \mathbf{u}_{i|k}$ denotes the predicted state, output, input vectors at time instant $i$ computed at time instant $k$. The above equations for $\mathbf{X}_{k+1}$ and $\mathbf{Y}_{k+1}$ are the central equations in output and state-based LMPC as discussed next.
\subsubsection{Output-based LMPC}

Output-based MPC uses the predicted output and control input sequence to construct the cost function. The objective is to find the optimal control sequence to regulate or track the output towards its reference. The output reference is denoted by $\mathbf{y}_{\mathrm{r}}$, which can be a constant (set point tracking or regulation problem) or time-varying (trajectory tracking problem). The following derivation is for regulation problems and can be extended to trajectory tracking problems by replacing $\mathbf{y}_{\mathrm{r}}$ with $\mathbf{y}_{\mathrm{r}_k}$. 
For a prediction horizon $\mathrm{N}$ and control horizon $\mathrm{N}_{\text{c}}$, the cost function for an output-based MPC at time instant $k$ is defined as:
\begin{equation}
\label{eqmpcjknc}
    J_{y}=\sum_{j=k+1}^{k+\mathrm{N}}[\mathbf{y}_{\mathrm{r}}-\mathbf{y}_{j|k}]^{\top}\mathbf{Q}_{\mathbf{y}_{j|k}}[\mathbf{y}_{\mathrm{r}}-\mathbf{y}_{j|k}]+\sum_{i=k}^{k+\mathrm{N}_{\text{c}}-1}[\mathbf{u}_{\mathrm{r}}-\mathbf{u}_{i|k}]^{\top}\mathbf{R}_{\mathbf{u}_{i|k}}[\mathbf{u}_{\mathrm{r}}-\mathbf{u}_{i|k}] 
\end{equation}
where  $\mathbf{Q}_{\mathbf{y}_{j|k}}\in \mathbb{R}^{p \times p}, \mathbf{R}_{\mathbf{u}_{i|k}}\in \mathbb{R}^{m \times m}$ are the output and input weighting matrices,  respectively. In the rest of the paper, we consider $\mathrm{N}_{\text{c}}=\mathrm{N}$ which modifies the above cost function as:
\begin{equation}
\label{eqmpcjk}
    J_{y}=\sum_{i=k}^{k+\mathrm{N}-1}[\mathbf{y}_{\mathrm{r}}-\mathbf{y}_{i+1|k}]^{\top}\mathbf{Q}_{\mathbf{y}_{i+1|k}}[\mathbf{y}_{\mathrm{r}}-\mathbf{y}_{i+1|k}]+[\mathbf{u}_{\mathrm{r}}-\mathbf{u}_{i|k}]^{\top}\mathbf{R}_{\mathbf{u}_{i|k}}[\mathbf{u}_{\mathrm{r}}-\mathbf{u}_{i|k}] 
\end{equation}
using which the output-based LMPC problem is defined as follows:
\begin{problem}[Output-based LMPC]
\label{prolmpc}
Given the current output $\mathbf{y}_{k|k},$ compute the control input sequence $\mathbf{U}_{k}$
for the LTI system in Eq. (\ref{eqlin}) by solving:
\begin{equation}
\label{eqlmpc}
    \begin{aligned}
    \underset{\mathbf{U}_{k},\mathbf{Y}_{k+1}}{\inf} ~&~  J_{y}\\ 
   \mathrm{subject \hspace{0.1cm}to} ~&~  \mathbf{Y}_{k+1}\in \mathbb{Y}^{\mathrm{N}}, \hspace{.2cm} \mathbf{U}_{k} \in \mathbb{U}^{\mathrm{N}} \\
    ~&~ \mathbf{x}_{i+1|k}=\mathbf{A}\mathbf{x}_{i|k}+\mathbf{B}\mathbf{u}_{i|k}\\
    ~&~ \mathbf{y}_{i|k}=\mathbf{C}\mathbf{x}_{i|k}, \hspace{.7cm}k\in \mathbb{T}, i=k,...,k+\mathrm{N}-1.
    \end{aligned}
\end{equation}
\end{problem}
Using $\mathbf{Y}_{k+1}$, $\mathbf{U}_{k}$ and defining $\mathbf{Q}_{\mathbf{Y}} \in \mathbb{R}^{p\mathrm{N}\times p\mathrm{N}}$,  
 $\mathbf{R}_{\mathbf{U}} \in \mathbb{R}^{m\mathrm{N}\times m\mathrm{N}},$ $\mathbf{Y}_{\mathrm{r}} \in \mathbb{R}^{p\mathrm{N}}$, $\mathbf{U}_{\mathrm{r}} \in \mathbb{R}^{m\mathrm{N}}$ as follows:
\begin{equation}
\label{eqqyru}
{{
    \mathbf{Q}_{\mathbf{Y}}=\left[\begin{matrix}
\mathbf{Q}_{\mathbf{y}_{k+1|k}} & \dots &  \mathbf{0}\\ \vdots &  &\vdots  \\ \mathbf{0} & \dots & \mathbf{Q}_{\mathbf{y}_{k+\mathrm{N}|k}}  
\end{matrix}\right] \hspace{0.5cm} \mathbf{R}_{\mathbf{U}}=\left[\begin{matrix}
\mathbf{R}_{\mathbf{u}_{k|k}}  &\dots  &\mathbf{0} \\  \\\vdots &  &\vdots \\ \mathbf{0} & \dots & \mathbf{R}_{\mathbf{u}_{k+\mathrm{N}-1|k}}
\end{matrix}\right] \hspace{0.5cm} \mathbf{Y}_{\mathrm{r}} = \left[\begin{matrix}
\mathbf{y}_{\mathrm{r}}  \\  \vdots \\ \mathbf{y}_{\mathrm{r}}
\end{matrix}\right] \hspace{0.5cm} \mathbf{U}_{\mathrm{r}} =  \left[\begin{matrix}
\mathbf{u}_{\mathrm{r}}  \\  \vdots \\ \mathbf{u}_{\mathrm{r}}
\end{matrix}\right] }}
\end{equation}
the MPC optimization problem in Eq. (\ref{eqlmpc}) can be rewritten as:
\begin{equation}
\label{eqoptympc2}
\begin{aligned}
 \underset{\mathbf{U}_{k},\mathbf{Y}_{k+1}}{\inf}  ~&~  [\mathbf{Y}_{r}-\mathbf{Y}_{k+1}]^{\top}\mathbf{Q}_{\mathbf{Y}} [\mathbf{Y}_{r}-\mathbf{Y}_{k+1}] + [\mathbf{U}_{r}-\mathbf{U}_{k}]^{\top} \mathbf{R}_{\mathbf{U}}[\mathbf{U}_{r}-\mathbf{U}_{k}] \\
\mathrm{subject \hspace{0.1cm}to} ~&~  \mathbf{Y}_{k+1} \in \mathbb{Y}^{\mathrm{N}}, \hspace{0.2cm}
\mathbf{U}_{k} \in \mathbb{U}^{\mathrm{N}}\\
~&~  \mathbf{Y}_{k+1}=\mathbf{A}_{\mathbf{Y}} \mathbf{x}_{k|k}+\mathbf{B}_{\mathbf{Y}} \mathbf{U}_{k}.
     \end{aligned}
\end{equation}
Substituting $\mathbf{Y}_{k+1}$ from Eq. (\ref{eqypred}) in Eq. (\ref{eqoptympc2}),
the MPC optimization problem can be represented with only $\mathbf{U}_{k}$ as the decision variable:
\begin{equation}
\label{eqoptympc}
\begin{aligned}
 \underset{\mathbf{U}_{k}}{\inf}  ~&~  [\mathbf{Y}_{r}-(\mathbf{A}_{\mathbf{Y}} \mathbf{x}_{k|k}+\mathbf{B}_{\mathbf{Y}} \mathbf{U}_{k})]^{\top}\mathbf{Q}_{\mathbf{Y}} [\mathbf{Y}_{r}-(\mathbf{A}_{\mathbf{Y}} \mathbf{x}_{k|k}+\mathbf{B}_{\mathbf{Y}} \mathbf{U}_{k})] + [\mathbf{U}_{r}-\mathbf{U}_{k}]^{\top} \mathbf{R}_{\mathbf{U}}[\mathbf{U}_{r}-\mathbf{U}_{k}] \\
\mathrm{subject \hspace{0.1cm}to} ~&~ \mathbf{A}_{\mathbf{Y}} \mathbf{x}_{k|k}+\mathbf{B}_{\mathbf{Y}} \mathbf{U}_{k} \in \mathbb{Y}^{\mathrm{N}}, \hspace{0.2cm}
\mathbf{U}_{k} \in \mathbb{U}^{\mathrm{N}}.
     \end{aligned}
\end{equation}
In the case of LMPC, the optimization problem can be represented as a QP. This can be achieved by expanding the
cost function in Eq. (\ref{eqoptympc})  and discarding the terms independent of $\mathbf{U}_{k}$:
\begin{equation}
\label{eqjkyu2}
    \begin{aligned}
      J_{y} & = (\mathbf{A}_{\mathbf{Y}}\mathbf{x}_{k|k}+\mathbf{B}_{\mathbf{Y}}\mathbf{U}_{k})^{\top}\mathbf{Q}_{\mathbf{Y}} (\mathbf{A}_{\mathbf{Y}}\mathbf{x}_{k|k}+\mathbf{B}_{\mathbf{Y}}\mathbf{U}_{k}) + \mathbf{U}_{k}^{\top} \mathbf{Q}_{\mathbf{U}} \mathbf{U}_{k}-2(\mathbf{A}_{\mathbf{Y}}\mathbf{x}_{k|k}+\mathbf{B}_{\mathbf{Y}}\mathbf{U}_{k})^{\top} \mathbf{Q}_{\mathbf{Y}} \mathbf{Y}_{\mathrm{r}}-2\mathbf{U}_{k}^{\top} \mathbf{Q}_{\mathbf{U}} \mathbf{U}_{\mathrm{r}}\\
      &=\mathbf{U}_{k}^{\top}  \underbrace{ [\mathbf{B}_{\mathbf{Y}}^{\top}\mathbf{Q}_{\mathbf{Y}} \mathbf{B}_{\mathbf{Y}} +\mathbf{Q}_{\mathbf{U}}]}_{ \mathbf{H}_{\mathbf{Y}} }  \mathbf{U}_{k}  +   \mathbf{U}_{k}^{\top} \underbrace{[2\mathbf{B}_{\mathbf{Y}}^{\top}\mathbf{Q}_{\mathbf{Y}} (\mathbf{A}_{\mathbf{Y}}\mathbf{x}_{k|k} - \mathbf{Y}_{\mathrm{r}})  - 2\mathbf{Q}_{\mathbf{U}} \mathbf{U}_{\mathrm{r}} ]}_{\mathbf{q}_{\mathbf{Y}} } + \underbrace{ \mathbf{x}_{k|k}^{\top}\mathbf{A}_{\mathbf{Y}}^{\top}\mathbf{Q}_{\mathbf{Y}} [\mathbf{A}_{\mathbf{Y}}\mathbf{x}_{k|k} -2\mathbf{Y}_{\mathrm{r}}]  }_{\text{Terms that does not depend on } \mathbf{U}_{k} } \\
      &=\mathbf{U}_{k}^{\top} \mathbf{H}_{\mathbf{Y}} \mathbf{U}_{k} + \mathbf{U}_{k}^{\top} \mathbf{q}_{\mathbf{Y}}
    \end{aligned}
\end{equation}
where $\mathbf{H}_{\mathbf{Y}} \in \mathbb{R}^{m\mathrm{N}\times m\mathrm{N}}$ and $\mathbf{q}_{\mathbf{Y}} \in \mathbb{R}^{m\mathrm{N}}$. Similarly, Using $\mathbf{Y}_{k+1}$  and $\mathbf{U}_{k}$ from Eq. (\ref{eqypred}), the constraints can be represented as:
    \begin{align}
    \label{eqfy}
  &\underbrace{\left[\begin{matrix}
\mathbf{F}_{\mathbf{y}} &\mathbf{0} & \dots & \mathbf{0} \\\mathbf{0} & \mathbf{F}_{\mathbf{y}} & \dots & \mathbf{0} \\ \vdots &\vdots  &  & \vdots \\\mathbf{0} & \mathbf{0} & \dots & \mathbf{F}_{\mathbf{y}}
\end{matrix}\right]}_{\mathbf{F}_{\mathbf{Y}}}\underbrace{\left[\begin{matrix}
\mathbf{y}_{k+1|k}\\\mathbf{y}_{k+2|k}\\\vdots\\ \mathbf{y}_{k+\mathrm{N}|k}
\end{matrix}\right]}_{\mathbf{Y}_{k+1}} \leq \underbrace{\left[\begin{matrix}
\mathbf{g}_{\mathbf{y}}\\\mathbf{g}_{\mathbf{y}} \\\vdots\\\mathbf{g}_{\mathbf{y}}
\end{matrix}\right]}_{\mathbf{g}_{\mathbf{Y}}} \\
\label{eqfu}
&\underbrace{\left[\begin{matrix}
\mathbf{F}_{\mathbf{u}} & \mathbf{0} & \dots & \mathbf{0}\\ \mathbf{0}& \mathbf{F}_{\mathbf{u}} & \dots & \mathbf{0} \\\vdots &\vdots  & & \vdots \\ \mathbf{0}& \mathbf{0}& \dots & \mathbf{F}_{\mathbf{u}}
\end{matrix}\right]}_{\mathbf{F}_{\mathbf{U}}}
   \underbrace{\left[\begin{matrix}
\mathbf{u}_{k|k}\\\mathbf{u}_{k+1|k}\\\vdots\\ \mathbf{u}_{k+\mathrm{N}-1|k}
\end{matrix}\right]}_{\mathbf{U}_{k}} \leq \underbrace{\left[\begin{matrix}
\mathbf{g}_{\mathbf{u}}\\\mathbf{g}_{\mathbf{u}} \\\vdots\\\mathbf{g}_{\mathbf{u}}
\end{matrix}\right]}_{\mathbf{g}_{\mathbf{U}}}  
    \end{align}
where $\mathbf{F}_{\mathbf{Y}} \in \mathbb{R}^{c_{y}\mathrm{N}\times p\mathrm{N}}$, $\mathbf{g}_{\mathbf{Y}} \in \mathbb{R}^{c_{y}\mathrm{N}}$, $\mathbf{F}_{\mathbf{U}} \in \mathbb{R}^{c_{u}\mathrm{N}\times m\mathrm{N}}$ and $\mathbf{g}_{\mathbf{U}} \in \mathbb{R}^{c_{u}\mathrm{N}}$.
Substituting $\mathbf{Y}_{k+1}$ from Eq. (\ref{eqypred}) in Eq. (\ref{eqfy}) and combining the constraints gives:
   \begin{equation}
    \label{eqfz2} \underbrace{\left[\begin{matrix}
\mathbf{F}_{\mathbf{Y}}\mathbf{B}_{\mathbf{Y}}  \\ \mathbf{F}_{\mathbf{U}}
\end{matrix}\right]}_{\mathbf{F}_{\mathbf{YU}}} \mathbf{U}_{k} \leq  \underbrace{\left[\begin{matrix}
\mathbf{g}_{\mathbf{Y}}-\mathbf{F}_{\mathbf{Y}} \mathbf{A}_{\mathbf{Y}}\mathbf{x}_{k|k} \\ \mathbf{g}_{\mathbf{U}}
\end{matrix}\right]}_{\mathbf{g}_{\mathbf{YU}}}
   \end{equation}
where $\mathbf{F}_{\mathbf{YU}} \in \mathbb{R}^{(c_{y}+c_{u})\mathrm{N}\times m\mathrm{N}}$ and $\mathbf{g}_{\mathbf{YU}}\in \mathbb{R}^{c_{y}+c_{u}}.$  Using the cost function in Eq. (\ref{eqjkyu2}) and constraint in Eq. (\ref{eqfz2}), the optimization problem for output-based LMPC can be represented as:  
\begin{equation}
\label{eqmpcqp}
\begin{aligned}
 \underset{\mathbf{U}_{k}}{\inf} \hspace{.2cm}&     \mathbf{U}_{k}^{\top}\mathbf{H}_{\mathbf{Y}}\mathbf{U}_{k} +  \mathbf{U}_{k}^{\top}\mathbf{q}_{\mathbf{Y}}\\
 \mathrm{subject \hspace{0.1cm}to} ~&~ \mathbf{F}_{\mathbf{YU}}\mathbf{U}_{k} \leq \mathbf{g}_{\mathbf{YU}}
     \end{aligned}
\end{equation}
which is a QP as in Eq. (\ref{eqqp}) with $\mathbf{U}_{k}$ as the decision vector. The solution of this QP gives the optimal control sequence $\mathbf{U}_{k}^{*}.$ 
This results in the
MPC control law: 
\begin{equation}
\label{eqmpcuk}
    \mathbf{u}_{k|k}=\mathbf{U}_{k_{(1:m)}}^{*}.
\end{equation}

Here, one of the challenges associated with the implementation of output-based MPC is the requirement of the current state $\mathbf{x}_{k|k}$ (which is normally unknown) in computing the term $\mathbf{q}_{\mathbf{Y}}$ in the cost function $J_{y}$. Therefore, we need to estimate the current state from the input-output data for which state estimators such as Kalman filters can be used. Next, we move on to the state-based LMPC.

\subsubsection{State-based LMPC}
State-based MPC uses the predicted state and control input sequence to construct the cost function. The state reference is denoted by $\mathbf{x}_{\mathrm{r}}$, which can be a constant or time-varying. For a prediction horizon $\mathrm{N}$, the cost function for the state-based LMPC at time instant $k$ is defined as
\begin{equation}
\label{eqmpcjkx}
    J_x=\sum_{i=k}^{k+\mathrm{N}-1}[\mathbf{x}_{\mathrm{r}}-\mathbf{x}_{i+1|k}]^{\top}\mathbf{Q}_{\mathbf{x}_{i+1|k}}[\mathbf{x}_{\mathrm{r}}-\mathbf{x}_{i+1|k}]+[\mathbf{u}_{\mathrm{r}}-\mathbf{u}_{i|k}]^{\top}\mathbf{R}_{\mathbf{u}_{i|k}}[\mathbf{u}_{\mathrm{r}}-\mathbf{u}_{i|k}] 
\end{equation}
where $\mathbf{Q}_{\mathbf{x}_{i+1|k}}\in \mathbb{R}^{n \times n}$ is the state weighting matrix, $\mathbf{R}_{\mathbf{u}_{i|k}}\in \mathbb{R}^{m \times m}$ is the input weighting matrix. 
Using the above cost function and constraints in Eq. (\ref{eqconstr}),
the state-based LMPC problem is defined as below:
\begin{problem}[State-based LMPC]
\label{prslmpc}
Given the current state $\mathbf{x}_{k|k}$, compute the control input sequence $\mathbf{U}_{k}$
for the LTI system in Eq. (\ref{eqlin}) by solving:
\begin{equation}
\label{eqlmpcx}
    \begin{aligned}
    \underset{\mathbf{U}_{k},\mathbf{X}_{k+1}}{\inf} ~&~  J_{x}\\ 
   \mathrm{subject \hspace{0.1cm}to} ~&~ \mathbf{X}_{k+1}\in \mathbb{X}^{\mathrm{N}},\hspace{.2cm} \mathbf{U}_{k} \in \mathbb{U}^{\mathrm{N}} \\
    ~&~ \mathbf{x}_{i+1|k}=\mathbf{A}\mathbf{x}_{i|k}+\mathbf{B}\mathbf{u}_{i|k}, \hspace{.7cm}k\in \mathbb{T}, i=k,...,k+\mathrm{N}-1.
    \end{aligned}
\end{equation}
\end{problem}
Using $\mathbf{X}_{k+1}$, $\mathbf{U}_{k}$ from Eq. (\ref{eqxpred}) and defining $\mathbf{Q}_{\mathbf{X}} \in \mathbb{R}^{n\mathrm{N}\times n\mathrm{N}}$,  
 $\mathbf{X}_{\mathrm{r}} \in \mathbb{R}^{n\mathrm{N}}$ similar to Eq. (\ref{eqqyru}), the MPC optimization problem in Eq. (\ref{eqlmpcx}) can be represented as:
\begin{equation}
\label{eqoptxmpc}
\begin{aligned}
 \underset{\mathbf{U}_{k}}{\inf}  ~&~  [\mathbf{X}_{r}-(\mathbf{A}_{\mathbf{X}} \mathbf{x}_{k|k}+\mathbf{B}_{\mathbf{X}} \mathbf{U}_{k})]^{\top}\mathbf{Q}_{\mathbf{X}} [\mathbf{X}_{r}-(\mathbf{A}_{\mathbf{X}} \mathbf{x}_{k|k}+\mathbf{B}_{\mathbf{X}} \mathbf{U}_{k})] + [\mathbf{U}_{r}-\mathbf{U}_{k}]^{\top} \mathbf{R}_{\mathbf{U}}[\mathbf{U}_{r}-\mathbf{U}_{k}] \\
\mathrm{subject \hspace{0.1cm}to} ~&~ \mathbf{A}_{\mathbf{X}} \mathbf{x}_{k|k}+\mathbf{B}_{\mathbf{X}} \mathbf{U}_{k} \in \mathbb{X}^{\mathrm{N}}, \hspace{0.2cm}
\mathbf{U}_{k} \in \mathbb{U}^{\mathrm{N}}.
     \end{aligned}
\end{equation}
 Similar to the output-based LMPC in Eq. ({\ref{eqmpcqp}}), the optimization problem for state-based LMPC can be represented as:  
\begin{equation}
\label{eqmpcqpx}
\begin{aligned}
 \underset{\mathbf{U}_{k}}{\inf} \hspace{.2cm}&     \mathbf{U}_{k}^{\top}\mathbf{H}_{\mathbf{X}}\mathbf{U}_{k} +  \mathbf{U}_{k}^{\top}\mathbf{q}_{\mathbf{X}}\\
 \mathrm{subject \hspace{0.1cm}to} ~&~ \mathbf{F}_{\mathbf{XU}}\mathbf{U}_{k} \leq \mathbf{g}_{\mathbf{XU}}
     \end{aligned}
\end{equation}
where $\mathbf{H}_{\mathbf{X}}=\mathbf{B}_{\mathbf{X}}^{\top}\mathbf{Q}_{\mathbf{X}} \mathbf{B}_{\mathbf{X}} +\mathbf{Q}_{\mathbf{U}} \in \mathbb{R}^{m\mathrm{N}\times m\mathrm{N}},$  $\mathbf{q}_{\mathbf{X}}= 2\mathbf{B}_{\mathbf{X}}^{\top}\mathbf{Q}_{\mathbf{X}} (\mathbf{A}_{\mathbf{X}}\mathbf{x}_{k|k} - \mathbf{X}_{\mathrm{r}})  - 2\mathbf{Q}_{\mathbf{U}} \mathbf{U}_{\mathrm{r}} \in \mathbb{R}^{m\mathrm{N}}$ and $\mathbf{F}_{\mathbf{XU}},\mathbf{g}_{\mathbf{XU}}$ are derived similar to Eq. (\ref{eqfz2}). The solution of this QP gives  $\mathbf{U}_{k}^{*} $, from which the first element is applied to the system as in Eq. (\ref{eqmpcuk}).
Next, we move on to the output and state-based NMPC design.

\subsection{Conventional NMPC}
Conventional NMPC uses a nonlinear model (in state-space form) of the system for designing MPC control law where the cost function is quadratic as in Eq. (\ref{eqmpcjk}) or (\ref{eqmpcjkx}). 
 Consider a discrete-time nonlinear system:
\begin{equation}
    \label{eqnlfx} 
    \begin{aligned} \mathbf{x}_{k+1}&=\mathbf{f}(\mathbf{x}_{k},\mathbf{u}_{k})\\
    \mathbf{y}_{k}&=\mathbf{h}(\mathbf{x}_{k})
    \end{aligned}
\end{equation}
where $k\in \mathbb{T}$, $\mathbf{x}_{k}\in \mathbb{X} \subseteq \mathbb{R}^{n}$ ,  $\mathbf{u}_{k}\in \mathbb{U} \subseteq \mathbb{R}^{m}$, $\mathbf{f}:\mathbb{X} \times \mathbb{U} \rightarrow \mathbb{X}$ is the state function that maps the current state $\mathbf{x}_{k}$ to the next state $\mathbf{x}_{k+1}$ for the control action $\mathbf{u}_{k},$ and $\mathbf{h}:\mathbb{X}  \rightarrow \mathbb{Y}$ is the output function.
The input, state, and output constraint sets are defined as in (\ref{eqconstr}). For a given initial condition $\mathbf{x}_{k|k}$ and control sequence $\mathbf{U}_{k}$, the predicted state and output sequence can be computed as:
\begin{align}
 \label{eqxprednl}
&\mathbf{X}_{k+1}=\left[\begin{matrix} \mathbf{x}_{k+1|k} \\  \vdots \\ \mathbf{x}_{k+\mathrm{N}|k}\end{matrix}\right]=  \left[\begin{matrix} \mathbf{f}(\mathbf{x}_{k|k},\mathbf{u}_{k|k})  \\  
 \vdots \\  
\mathbf{f}(\dots\mathbf{f}(\mathbf{x}_{k|k},\mathbf{u}_{k|k}), \mathbf{u}_{k+\mathrm{N}-1|k})
\end{matrix}\right]=\mathbf{f}_{\mathrm{xp}}(\mathbf{x}_{k|k},\mathbf{U}_{k})\\
\label{eqyprednl}
&\mathbf{Y}_{k+1}=\left[\begin{matrix} \mathbf{y}_{k+1|k} \\  \vdots \\ \mathbf{y}_{k+\mathrm{N}|k}\end{matrix}\right]=\left[\begin{matrix} \mathbf{h}(\mathbf{x}_{k+1|k}) \\  \vdots \\ \mathbf{h}(\mathbf{x}_{k+\mathrm{N}|k})\end{matrix}\right]= \mathbf{f}_{\mathrm{yp}}(\mathbf{x}_{k|k},\mathbf{U}_{k})
\end{align}
which are used to construct model-based equality constraints in output-based and state-based NMPC as discussed next.
\subsubsection{Output-based NMPC}
Output-based NMPC uses the predicted output and control input sequence to construct the cost function. However, unlike LMPC, the model used for prediction is nonlinear in the case of NMPC.
The cost function for the output-based NMPC at time instant $k$ is defined as
in Eq. (\ref{eqmpcjk}), using which the output-based NMPC problem can be defined as:
\begin{problem}[Output-based NMPC]
\label{pronmpc}
For the nonlinear system (\ref{eqnlfx}) with the output $\mathbf{y}_{k|k}$ given,
compute the control input sequence $\mathbf{U}_{k}$ by solving:
\begin{equation}
\label{eqnmpc}
    \begin{aligned}    \underset{\mathbf{U}_{k},\mathbf{Y}_{k+1}}{\inf} \hspace{.2cm} &J_{y}\\ \mathrm{subject \hspace{0.1cm}to}
    ~&~  \mathbf{Y}_{k+1}\in \mathbb{Y}^{\mathrm{N}},\hspace{0.2cm} \mathbf{U}_{k} \in \mathbb{U}^{\mathrm{N}}\\
    ~&~ \mathbf{x}_{i+1|k}=\mathbf{f}(\mathbf{x}_{i|k},\mathbf{u}_{i|k}) \\
    ~&~\mathbf{y}_{i|k}=\mathbf{h}(\mathbf{x}_{i|k}), \hspace{0.7cm}k\in \mathbb{T},i=k,...,k+\mathrm{N}-1.
    \end{aligned}
\end{equation}
\end{problem}

The predicted output sequence $\mathbf{Y}_{k+1}$ and control sequence $\mathbf{U}_{k}$ in Eq. (\ref{eqyprednl}) can be used to 
rewrite the optimization problem for output-based NMPC as:
\begin{equation}
\label{eqnmpc2y}
\begin{aligned}
 \underset{\mathbf{U}_{k}}{\inf}  ~&~  [\mathbf{Y}_{r}-\mathbf{f}_{\mathrm{yp}}(\mathbf{x}_{k|k},\mathbf{U}_{k})]^{\top}\mathbf{Q}_{\mathbf{Y}} [\mathbf{Y}_{r}-\mathbf{f}_{\mathrm{yp}}(\mathbf{x}_{k|k},\mathbf{U}_{k})] + [\mathbf{U}_{r}-\mathbf{U}_{k}]^{\top} \mathbf{R}_{\mathbf{U}}[\mathbf{U}_{r}-\mathbf{U}_{k}] \\
\mathrm{subject \hspace{0.1cm}to} ~&~ \mathbf{f}_{\mathrm{yp}}(\mathbf{x}_{k|k},\mathbf{U}_{k}) \in \mathbb{Y}^{\mathrm{N}}, \hspace{0.2cm}
\mathbf{U}_{k} \in \mathbb{U}^{\mathrm{N}}.
     \end{aligned}
\end{equation}
Here the cost function and equality constraint contains the nonlinear function $\mathbf{f}_{\mathrm{yp}}$ (built using the nonlinear functions $\mathbf{f}$ and $\mathbf{h}$) which makes Eq. (\ref{eqnmpc2y}) an NLP.
The output-based NMPC scheme is obtained by solving this NLP at every time instant and applying the first element of $\mathbf{U}_{k}^{*}$ as the control input.

\subsubsection{State-based NMPC}
State-based NMPC uses the predicted state and control input sequence to construct the cost function, where the states are predicted using a nonlinear model.
The cost function for the state-based NMPC is defined as
in Eq. (\ref{eqmpcjkx}), using which the state-based NMPC problem can be defined as:
\begin{problem}[State-based NMPC]
\label{prsnmpc}
For the nonlinear system (\ref{eqnlfx}) with the current state $\mathbf{x}_{k|k}$ given,
compute the control input sequence $\mathbf{U}_{k}$ by solving:
\begin{equation}
\label{eqnmpcx}
    \begin{aligned}
    \underset{\mathbf{U}_{k},\mathbf{X}_{k+1}}{\inf} \hspace{.2cm} &J_{y}\\ \mathrm{subject \hspace{0.1cm}to}
    ~&~\mathbf{X}_{k+1}\in \mathbb{X}^{\mathrm{N}},\hspace{0.2cm} \mathbf{U}_{k} \in \mathbb{U}^{\mathrm{N}}\\
    ~&~ \mathbf{x}_{i+1|k}=\mathbf{f}(\mathbf{x}_{i|k},\mathbf{u}_{i|k}), \hspace{0.7cm}i=k,...,k+\mathrm{N}-1.
    \end{aligned}
\end{equation}
\end{problem}
The predicted state and control sequences in Eq. (\ref{eqxprednl}) can be used to represent the optimization problem for state-based NMPC as:
\begin{equation}
\label{eqnmpc2x}
\begin{aligned}
 \underset{\mathbf{U}_{k}}{\inf}  ~&~  [\mathbf{X}_{r}-\mathbf{f}_{\mathrm{xp}}(\mathbf{x}_{k|k},\mathbf{U}_{k})]^{\top}\mathbf{Q}_{\mathbf{X}} [\mathbf{X}_{r}-\mathbf{f}_{\mathrm{xp}}(\mathbf{x}_{k|k},\mathbf{U}_{k})] + [\mathbf{U}_{r}-\mathbf{U}_{k}]^{\top} \mathbf{R}_{\mathbf{U}}[\mathbf{U}_{r}-\mathbf{U}_{k}] \\
\mathrm{subject \hspace{0.1cm}to} ~&~ \mathbf{f}_{\mathrm{xp}}(\mathbf{x}_{k|k},\mathbf{U}_{k}) \in \mathbb{X}^{\mathrm{N}}, \hspace{0.2cm}
\mathbf{U}_{k} \in \mathbb{U}^{\mathrm{N}}.
     \end{aligned}
\end{equation}
The state-based NMPC scheme is obtained by solving this NLP at every time instant and applying the first element of $\mathbf{U}_{k}^{*}$ as the control input. Numerical examples that illustrate the conventional LMPC and NMPC can be found in \cite{bMA1}. It is important to note that the control reference $\mathbf{U}_{\text{r}}$ is generally unknown in practical applications. It can, however, be computed from the output or state reference using steady-state conditions, or alternatively, it may be selected as the optimal control sequence obtained at the previous time step. 
\par The equality constraints in optimization problems (Problem \ref{prolmpc} to Problem \ref{prsnmpc}) can be called as \textbf{model-based constraints}, and to define them, we need the model information. In D-MPC the model information is assumed to be unknown and needed to be extracted from data.
The rest of the paper discusses various approaches in model-based D-MPC, which primarily focuses on the identification of deterministic state-space models in discrete-time. The decision to limit the scope to this specific category is motivated by the fact that both conventional MPC and D-MPC are predominantly designed using state-space models. Next, we will discuss D-LMPC, which uses linear state-space models identified using input-output data for MPC design.
\section{Data-driven Linear MPC}

Data-driven Linear MPC (D-LMPC) uses input-output data to either identify a linear model as in Eq. (\ref{eqlin}) or  an equivalent set of constraints that represent the system's predictive behavior, i.e., predicted output or state as a function of control input.  The available information  consists of the input and output data:
   \begin{equation}
      \label{equy}  
     \begin{aligned}
     &\mathbf{U}_{\mathrm{D}}=\left[\begin{matrix} \mathbf{u}_{0} & \mathbf{u}_{1} & \dots & \mathbf{u}_{{\mathrm{D}-1}}\end{matrix}\right]\\
     &\mathbf{Y}_{\mathrm{D}}=\left[\begin{matrix} \mathbf{y}_{1} & \mathbf{y}_{2} & \dots & \mathbf{y}_{{\mathrm{D}}}\end{matrix}\right]
        \end{aligned}  
    \end{equation}  
where $\mathrm{D}$ is the number of samples in the dataset. In order to distinguish the predicted output and control input sequence used in MPC, the training input and output sequences are denoted by $\mathbf{U}_{\mathrm{D}}$ and $\mathbf{Y}_{\mathrm{D}}$ which are already available or historical data.  In data-driven LMPC, the system parameters $\mathbf{A},\mathbf{B},\mathbf{C}$ are considered to be unknown. Instead, input-output data: $\mathbf{U}_{\mathrm{D}},\mathbf{Y}_{\mathrm{D}}$ is available, which can be used for identifying the system parameters and designing MPC control schemes. 
This leads to the following two directions of D-LMPC:
\begin{enumerate}
    \item \textbf{Model-based D-LMPC}: in which the training data is used for identifying a linear model, which is then used for solving the MPC problem. Examples:  realization method and prediction error method based LMPC.

    \item \textbf{Model-free D-LMPC}:  which directly solves the LMPC problem (without identifying a model first)
in which the model-based constraints are replaced by data-based constraints. Example: data-enabled predictive control.
\end{enumerate}
 This section focuses on approaches in model-based D-LMPC, i.e., SysID followed by MPC. A critical step in most SysID and MPC strategies involves expressing the predicted output or state trajectory as a function of the control input sequence. This is where the model comes into the picture: it is employed to establish a mathematical relationship between the predicted system output (or states) and the corresponding control input sequence. Through this relationship, the model helps in optimizing the control inputs within the MPC framework, ensuring that the predicted system behavior aligns with the desired performance objectives. In this section, we consider the identification of linear state-space models from input-output data, which is then used for designing MPC schemes.
 The linear state-space identification methods proposed in the past consist of:
    \begin{enumerate}
        \item \textbf{Realization methods} \cite{bBR66}-\cite{bWB99}: in which the system parameters are estimated from the training data by constructing the input and output Hankel matrices, followed by a factorization step such as singular value decomposition (SVD). Examples are the Ho-Kalman algorithm, Subspace-based methods, etc.
        \item \textbf{Prediction error methods} (PEMs) \cite{bRD17}: are optimization-based methods in which the model parameters are computed by minimizing a loss function of the output prediction error.        
    \end{enumerate}
In this paper, we will be mostly using PEMs for SysID, which can be used for the identification of both linear and nonlinear systems.
Before moving on to PEM-based identification, next we will discuss D-LMPC based on realization methods with numerical implementation.

\subsection{Ho-Kalman-Kung Algorithm based D-LMPC}
One of the initial approaches for linear state space identification is the Ho-Kalman algorithm, which is a realization-based method \cite{bBR66}. Ho-Kalman algorithm uses the discrete \textbf{impulse response} data for identifying the system parameters: $\mathbf{A},\mathbf{B},\mathbf{C}$ matrices. The discrete impulse response of a linear system refers to the system's output when subjected to a unit pulse input, often referred to as an impulse input. The Ho-Kalman-Kung algorithm is an extension of the Ho-Kalman algorithm proposed by Kung \cite{bSK78} which is based on the SVD-based factorization of the Hankel matrix constructed using the impulse response data. A \textbf{Hankel matrix} is characterized by constant entries along each ascending skew-diagonal. In other words, a Hankel matrix 
$\mathbf{H}$ is defined by its elements 
$h_{ij}$ , where the entry at position 
($i,j$) is constant for all pairs 
$i+j=c$, where 
$c$ is the skew-diagonal index. This structure is particularly useful in various system identification and signal processing tasks, as it captures the system's temporal or spatial dependencies in a compact form.
 Consider the  discrete time unit pulse function defined as:
\begin{equation}
    u_{k} =\begin{cases}
1, \hspace{0.3cm} & for \hspace{0.1cm} k=0\\
0, \hspace{0.3cm} &for \hspace{0.1cm} k=1,2,3,\dots
\end{cases}
\end{equation}
Applying this impulse input to a linear system with $\mathbf{x}_{0}=\mathbf{0}$ results in the unit-pulse response as:
\begin{equation}
\begin{aligned}  y_{1}&=\mathbf{C}\mathbf{x}_{1}=\mathbf{C}[\mathbf{A}\mathbf{x}_{0}+\mathbf{B}u_{0}]=\mathbf{CB}\\
    y_{2}&=\mathbf{C}\mathbf{x}_{2}=\mathbf{C}[\mathbf{A}^{2}\mathbf{x}_{0}+\mathbf{A}\mathbf{B}u_{0}+\mathbf{B}u_{1}]=\mathbf{C}\mathbf{A}\mathbf{B}\\
     & \hspace{0.2cm} \vdots\\
      y_{k} &= \mathbf{C}\mathbf{x}_{k}= \mathbf{C}\mathbf{A}^{k-1}\mathbf{B}.
    \end{aligned}
\end{equation}
 The unit pulse response terms are also known as Markov parameters
using which the Hankel matrix for the Ho-Kalman-Kung algorithm is defined as:
   \begin{equation}
   \label{eqhy1}
   \begin{aligned}
   \textbf{H}_{(\text{N},\text{H},y_{1})} &=\left[\begin{matrix} y_{1} & y_{2} &  \dots & y_{\text{H}}\\ y_{2} &  & & \vdots \\  \vdots & & &  \vdots \\ y_{\text{N}} & \dots   &  & y_{{\text{N}+\text{H}-1}}
\end{matrix}\right]=  \left[\begin{matrix} \textbf{CB} & \textbf{CAB}  & \dots & \textbf{C}\textbf{A}^{\text{H}-1}\textbf{B}\\ \textbf{CAB} &   & & \vdots \\ \vdots & & & \vdots  \\ \textbf{C}\textbf{A}^{\text{N}-1}\textbf{B} & \dots  & &  \textbf{C}\textbf{A}^{\text{N}+\text{H}-2}\textbf{B}
\end{matrix}\right]\\ &=   {{ \underbrace{\left[\begin{matrix}\textbf{C}\\\textbf{CA}\\\vdots\\\textbf{C}{\textbf{A}}^{\text{N}-1}
\end{matrix}\right]}_{\textbf{O}_{\text{N}}} \underbrace{ \left[\begin{matrix}\textbf{B}&\textbf{AB}&\dots&{\textbf{A}}^{\text{H}-1}\textbf{B}\end{matrix}\right]}_{\textbf{C}_{\text{H}}}  }}
\end{aligned}
   \end{equation}
   where $\mathbf{C}_{\mathrm{H}}$ and $\mathbf{O}_{\mathrm{N}}$ can be called as extended controllability and observability matrices where $\mathrm{N}\geq n$ and $\mathrm{H}\geq n.$ 
For $\mathrm{N}=\mathrm{H}=n$, the Hankel matrix becomes  the product of the observability and controllability matrix: $\mathbf{H}_{(\mathrm{n},\mathrm{n},y_{1})}=\mathbf{O}_{n}\mathbf{C}_{n}$.
   Now, the first row of $\mathbf{O}_{n}$ gives the $\mathbf{C}$ matrix and first column of $\mathbf{C}_{n}$ gives the $\mathbf{B}$ matrix. Further, the $\mathbf{A}$ matrix can be computed from  the relation $\mathbf{O}_{n_{(2:n,:)}}=\mathbf{O}_{n_{(1:n-1,:)}}\mathbf{A}$ and this results in the initial version of the Ho-Kalman algorithm.  However, in practice, the actual order of the system is unknown. In that case, the values of $\mathrm{N}$ and $\mathrm{H}$ are chosen sufficiently large and assume $\mathrm{N} \geq \mathrm{H} \geq n$. We have for controllable and observable systems, the rank of the controllability and observability matrix will be equal to $n$.  Therefor to find $n$, the Ho-Kalman-Kung algorithm uses SVD of the Hankel matrix in Eq. (\ref{eqhy1}) results in:
 \begin{equation}  
\label{eqhsvd}\mathbf{H}_{(\mathrm{N},\mathrm{H},y_{1})}= \mathbf{W}\left[\begin{matrix}\bm{\Sigma} & \mathbf{0}
\end{matrix}\right]_{(\mathrm{N}\times \mathrm{H})}\left[\begin{matrix}\mathbf{V}^{\top} \\ \mathbf{V}_{\mathrm{r}}^{\top}
\end{matrix}\right]_{(\mathrm{H}\times \mathrm{H})}=\mathbf{W}\bm{\Sigma}\mathbf{V}^{\top}=\mathbf{W}\bm{\Sigma}^{\frac{1}{2}}\bm{\Sigma}^{\frac{1}{2}}\mathbf{V}^{\top}=\underbrace{\mathbf{W}\bm{\Sigma}^{\frac{1}{2}} \mathbf{P}}_{\mathbf{O}_{\mathrm{N}}}\underbrace{\mathbf{P}^{-1}\bm{\Sigma}^{\frac{1}{2}}\mathbf{V}^{\top}}_{\mathbf{C}_{\mathrm{H}}}
 \end{equation}
where $\bm{\Sigma}=diag(\sigma_{1},\dots,\sigma_{\mathrm{N}})\in \mathbb{R}^{\mathrm{N}\times \mathrm{N}},$ $\mathbf{P}=\mathbb{R}^{\mathrm{N}\times \mathrm{N}}$ is any invertible matrix, $\mathbf{W}\in \mathbb{R}^{\mathrm{N}\times \mathrm{N}},$ and $\mathbf{V}\in \mathbb{R}^{\mathrm{H}\times \mathrm{N}}$. 
 The number of non-zero singular values (diagonal elements of $\bm{\Sigma} $) gives an approximation of the model order $n$ denoted by $s$. The resulting model with order $s$ is denoted by:
\begin{equation}
\label{eqlin}
\begin{aligned}
{ \mathbf{x}_{\mathrm{s}_{k+1}}}&=\mathbf{A}_{\mathrm{s}}\mathbf{x}_{\mathrm{s}_k}+\mathbf{B}_{\mathrm{s}}\mathbf{u}_{k}\\
\hat{\mathbf{y}}_{\mathrm{s}_k}&=\mathbf{C}_{\mathrm{s}}\mathbf{x}_{\mathrm{s}_k}
\end{aligned}
\end{equation}
where  $\mathbf{A}_{\mathrm{s}} \in \mathbb{R}^{s \times s},$ $\mathbf{B}_{\mathrm{s}} \in \mathbb{R}^{s \times m},$ $\mathbf{C}_{\mathrm{s}} \in \mathbb{R}^{p \times s},$ and $\mathbf{x}_{\mathrm{s}_k}\in \mathbb{R}^{s}$. 
 Note that, the singular values are nonnegative and ordered in decreasing magnitude. Consequently, the approximate model order $s$ can be computed by solving:
  \begin{equation}
\label{eqmo}
\begin{aligned}
s=\underset{i\in\{1,2,\dots,\mathrm{N}\}}{\mathrm{argmax}} \hspace{.2cm}  &i \\
\mathrm{subject \hspace{0.1cm} to}~&~ \sigma_{i} \geq \epsilon .
\end{aligned}
 \end{equation}
 where $\epsilon >0$ is a tolerance value. 
Using SVD, a rank$-s$ approximation to $\mathbf{H}_{(\mathrm{N},\mathrm{H},y_{1})}$ can be obtained as:
\begin{equation}
    \label{eqhy2}\mathbf{H}_{(\mathrm{N},\mathrm{H},y_{1})}= \underbrace{\mathbf{W}\bm{\Sigma}^{\frac{1}{2}} \mathbf{P}}_{\mathbf{O}_{\mathrm{N}}}\underbrace{\mathbf{P}^{-1}\bm{\Sigma}^{\frac{1}{2}}\mathbf{V}^{\top}}_{\mathbf{C}_{\mathrm{H}}} \approx \underbrace{\mathbf{W}_{\mathrm{s}}\bm{\Sigma}_{\mathrm{s}}^{\frac{1}{2}} \mathbf{P}_{\mathrm{s}}}_{\mathbf{O}_{\mathrm{N}_s}}\underbrace{\mathbf{P}_{\mathrm{s}}^{-1}\bm{\Sigma}_{\mathrm{s}}^{\frac{1}{2}}\mathbf{V}_{\mathrm{s}}^{\top}}_{\mathbf{C}_{\mathrm{H}_s}}
\end{equation}
where $\mathbf{W}_{\mathrm{s}}=\mathbf{W}_{(:,1:s)}$, $\mathbf{V}_{\mathrm{s}}=\mathbf{V}_{(:,1:s)}$, $\bm{\Sigma}_{\mathrm{s}}=\bm{\Sigma}_{(1:s,1:s)}$, $\mathbf{P}_{\mathrm{s}}=\mathbf{P}_{(1:s,1:s)},$ and  $\mathbf{O}_{\mathrm{N}_s},$  $\mathbf{C}_{\mathrm{H}_s} $ are  rank$-s$ approximations to extended controllability and observability matrices. Further, from the definition of $\mathbf{O}_{\mathbf{N}}$ in Eq. (\ref{eqhy1}) it can be observed that:
\begin{equation}
\label{eqnny2}
   \mathbf{O}_{{\mathrm{N}_s}_{(2:\mathrm{N},:)}}=   \mathbf{O}_{{\mathrm{N}_s}_{(1:\mathrm{N}-1,:)}}\mathbf{A}_{\mathrm{s}}
\end{equation} 
which can be solved for $s^{th}$ order system matrix. 
This gives the system parameters for an $s^{th}$ order state-space realization as:
\begin{equation}
\label{eqssest}
    \mathbf{A}_{\mathrm{s}}=  \mathbf{O}_{{\mathrm{N}_s}_{(1:\mathrm{N}-1,:)}}^{\dagger}  \mathbf{O}_{{\mathrm{N}_s}_{(2:\mathrm{N},:)}} \hspace{1cm} \mathbf{B}_{\mathrm{s}}=\mathbf{C}_{{\mathrm{H}_s}_{(:,1)}}  \hspace{1cm}\mathbf{C}_{\mathrm{s}}=\mathbf{O}_{{\mathrm{N}_s}_{(1,:)}}.
\end{equation}

Now, using the $\mathbf{A}_{\mathrm{s}},\mathbf{B}_{\mathrm{s}},\mathbf{C}_{\mathrm{s}}$ matrices, output or state-based LMPC can be designed as in Section \ref{seclmpc} with $\mathbf{A}=\mathbf{A}_{\mathrm{s}},$ $\mathbf{B}=\mathbf{B}_{\mathrm{s}}$, $\mathbf{C}=\mathbf{C}_{\mathrm{s}}$. The steps for the D-LMPC based on the Ho-Kalman-Kung algorithm are summarized below.
   
    \begin{algorithm}[H]
 
	\begin{algorithmic}[1] 
	
	\STATE Require the impulse response data $y_{0},y_{1},\dots,y_{\mathrm{D}}$.
    \STATE Select $\mathrm{N}$ and $\mathrm{H}$ such that $\mathrm{N}+\mathrm{H}-1 \leq \mathrm{D}$. 
		\STATE Compute $\mathbf{H}_{(\mathrm{N},\mathrm{H},y_{1})}$ as in Eq. (\ref{eqhy1}).
		\STATE Find SVD of $\mathbf{H}_{(\mathrm{N},\mathrm{H},y_{1})}$ as in Eq. (\ref{eqhsvd})
        \STATE Find an approximate model order $s$ as in Eq. (\ref{eqmo}).   
        \STATE Compute $\mathbf{O}_{\mathrm{N}_s}=$ and $\mathbf{C}_{\mathrm{H}_s}=$ as in Eq. (\ref{eqhy2}). 
        \STATE Use  $\mathbf{O}_{\mathrm{N}_s}$ and $\mathbf{C}_{\mathrm{H}_s}$ to find $\mathbf{A}_{\mathrm{s}}$, $\mathbf{B}_{\mathrm{s}}$, and $\mathbf{C}_{\mathrm{s}}$ as in Eq. (\ref{eqssest}).\\
        
        \STATE Use $\mathbf{A}_{\mathrm{s}}$, $\mathbf{B}_{\mathrm{s}}$, and $\mathbf{C}_{\mathrm{s}}$ to implement output or state based LMPC as in Section \ref{seclmpc}.
	\end{algorithmic}
	\caption{: Ho-Kalman-Kung based D-LMPC}
    \label{alghk}
\end{algorithm}

\subsection{D-LMPC based on Ho-Kalman-Kung algorithm: Numerical Example}
\begin{figure}[H] 
 		\begin{center}
 		\includegraphics [scale=.1625] {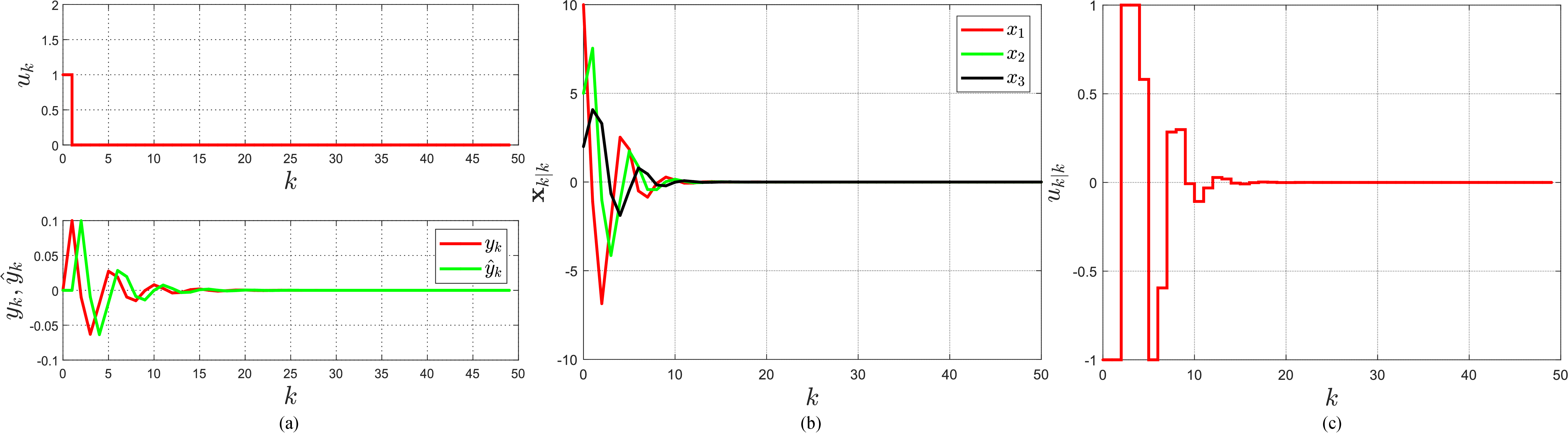}
 		\caption{{ Ho-Kalman-Kung based D-LMPC \hspace{.15cm} (a) Training input and output \hspace{.1cm}(b) State response  \hspace{.1cm}(c) Control input.}}	
        \label{fighok}
 	\end{center}
    
 \end{figure}

To illustrate the Ho-Kalman-Kung algorithm, we generate the impulse response data for a third-order LTI system for which the system parameters are given below:
\begin{equation}
    \mathbf{A}=\left[\begin{matrix} 0.2 &-0.4 & 0.5 \\0.7 & 0.3 & 0.6 \\ -0.5 & 0.1 & 0.6
\end{matrix}\right] \hspace{1cm} \mathbf{B}=\left[\begin{matrix} 0.1 \\0.2 \\ 0.1
\end{matrix}\right]  \hspace{1cm} \mathbf{C}=\left[\begin{matrix} 1 & 0 & 0
\end{matrix}\right] .
\end{equation}
The LTI system is simulated for $\mathrm{D}=50$ instants with a pulse input as $u_{k}.$ The generated impulse response data is used for identifying system parameters as in Algorithm  \ref{alghk} with $\mathrm{N}=5$ and $\mathrm{H}=5$. The identified system parameters are used for implementing D-LMPC for which
 the simulation parameters  are chosen as $\mathrm{N}_{\mathrm{T}}=50,$ $\mathrm{N}=10,$ $\mathbf{Q}_{\mathbf{x}}=\mathbf{I}_{3},\mathbf{R}_{\mathbf{u}}=0.5$ and $\mathbf{x}_{0}=\left[\begin{matrix} 10 & 5 & 2
\end{matrix}\right]^{T}.$ The constraint set is defined using the lower and upper bounds as $x_{min} \leq x_{i} \leq  x_{max},$ $i=1,2,3$ and $u_{min}\leq u \leq u_{max}$ with $x_{min}=-10,$ $x_{max}=10,$ $u_{min}=-1,$ $u_{max}=1$. The state reference is chosen as $\mathbf{x}_{\mathrm{r}}=\mathbf{0}$, which results in a stabilization problem.
     Fig. \ref{fighok}(a) shows the training (pulse) input and output (impulse response).  
     Fig. \ref{fighok}(b)
     shows the response of the LTI system with the Ho-Kalman-Kung based D-MPC. The response shows the states converge to the origin in less than 20 instnats.  The control input plot is given in Fig. \ref{fighok}(c), which shows that the control constraints are satisfied.

\subsection{Subspace Predictive Control}
Subspace predictive control (SPC) is a subspace-based D-LMPC approach introduced in \cite{bWB99}. SPC relies on training data as in Eq. (\ref{equy}) to construct input and output block Hankel matrices, which serve as a compact representation of the system's dynamic behavior. These Hankel matrices capture the system’s temporal relationships and are used to estimate the system's model parameters. By applying subspace identification techniques, the system's predicted output dynamics can be derived from the data, enabling the formulation of an MPC controller. 
In SPC, the past data is used for identifying a model-based constraint, and then the future data predicted using the model is used in MPC to compute the optimal control input. This leads to  the following horizons:
\begin{enumerate}
 \item \textbf{Future horizon} (prediction horizon) $\mathrm{N}$: number of future samples predicted using past data or current data.   
    \item \textbf{Past horizon} (measurement horizon) $\mathrm{M}$: number of lags of past data used for identifying the model.  
    \item \textbf{Training horizon} $\mathrm{H}$: decides the number of training samples used for identifying the model.
\end{enumerate}
In the SPC derivation next, we consider the past horizon equals future horizon $\mathrm{M}=\mathrm{N}$ and only uses $\mathrm{N}$, to simplify the notations. 
For the training data as in Eq. (\ref{equy}), and a given  N and H, the input and output Hankel matrices can be constructed as: 
\begin{equation}
\label{equpuf}
\begin{aligned}   &\mathbf{U}_{\mathrm{p}}=\left[\begin{matrix} \mathbf{u}_{0} &  \mathbf{u}_{1} & \dots & \mathbf{u}_{{\mathrm{H}-1}} \\ 
    \mathbf{u}_{1} &  \mathbf{u}_{2} & \dots & \mathbf{u}_{{\mathrm{H}}}\\
    \vdots & \vdots &  & \vdots\\
    \mathbf{u}_{{\mathrm{N}-1}} &  \mathbf{u}_{{\mathrm{N}}} & \dots & \mathbf{u}_{{\mathrm{N}+\mathrm{H}-2}}
\end{matrix}\right]=\left[\begin{matrix} \mathbf{U}_{0} & \mathbf{U}_{1} & \dots & \mathbf{U}_{{\mathrm{H}-1}}\end{matrix}\right] =\mathbf{H}_{(\mathrm{N},\mathrm{H},\mathbf{u}_{0})}\in \mathbb{R}^{m\mathrm{N}\times \mathrm{H}} \\ &\mathbf{U}_{\mathrm{f}}=\left[\begin{matrix} \mathbf{u}_{\mathrm{N}} &  \mathbf{u}_{{\mathrm{N}+1}} & \dots & \mathbf{u}_{{\mathrm{N}+\mathrm{H}-1}} \\ 
    \mathbf{u}_{{\mathrm{N}+1}} &  \mathbf{u}_{{\mathrm{N}+2}} & \dots & \mathbf{u}_{{\mathrm{N}+\mathrm{H}}}\\
    \vdots & \vdots &  & \vdots\\
    \mathbf{u}_{{2\mathrm{N}-1}} &  \mathbf{u}_{{2\mathrm{N}}} & \dots & \mathbf{u}_{{2\mathrm{N}+\mathrm{H}-2}}
\end{matrix}\right]=\left[\begin{matrix} \mathbf{U}_{\mathrm{N}} & \mathbf{U}_{{\mathrm{N}+1}} & \dots & \mathbf{U}_{{\mathrm{N}+\mathrm{H}-1}}\end{matrix}\right]=\mathbf{H}_{(\mathrm{N},\mathrm{H},\mathbf{u}_{\mathrm{N}})}  \in \mathbb{R}^{m\mathrm{N}\times \mathrm{H}}   \\
&\mathbf{Y}_{\mathrm{p}}=\left[\begin{matrix} \mathbf{y}_{1} &  \mathbf{y}_{2} & \dots & \mathbf{y}_{{\mathrm{H}}} \\ 
    \mathbf{y}_{2} &  \mathbf{y}_{3} & \dots & \mathbf{y}_{{\mathrm{H}+1}}\\
    \vdots & \vdots &  & \vdots\\
    \mathbf{y}_{{\mathrm{N}}} &  \mathbf{y}_{{\mathrm{N}+1}} & \dots & \mathbf{y}_{{\mathrm{N}+\mathrm{H}-1}}
\end{matrix}\right]=\left[\begin{matrix} \mathbf{Y}_{1} & \mathbf{Y}_{2} & \dots & \mathbf{Y}_{{\mathrm{H}}}\end{matrix}\right]=\mathbf{H}_{(\mathrm{N},\mathrm{H},\mathbf{y}_{1})} \in \mathbb{R}^{p\mathrm{N}\times \mathrm{H}}\\
    &\mathbf{Y}_{\mathrm{f}}=\left[\begin{matrix} \mathbf{y}_{{\mathrm{N}+1}} &  \mathbf{y}_{{\mathrm{N}+2}} & \dots & \mathbf{y}_{{\mathrm{N}+\mathrm{H}}} \\ 
    \mathbf{y}_{{\mathrm{N}+2}} &  \mathbf{y}_{{\mathrm{N}+3}} & \dots & \mathbf{y}_{{\mathrm{N}+\mathrm{H}+1}}\\
    \vdots & \vdots &  & \vdots\\
    \mathbf{y}_{{2\mathrm{N}}} &  \mathbf{y}_{{2\mathrm{N}+1}} & \dots & \mathbf{y}_{{2\mathrm{N}+\mathrm{H}-1}}
\end{matrix}\right]=\left[\begin{matrix} \mathbf{Y}_{{\mathrm{N}+1}} & \mathbf{Y}_{{\mathrm{N}+2}} & \dots & \mathbf{Y}_{{\mathrm{N}+\mathrm{H}}}\end{matrix}\right]=\mathbf{H}_{(\mathrm{N},\mathrm{H},\mathbf{y}_{{\mathrm{N}+1}})} \in \mathbb{R}^{p\mathrm{N}\times \mathrm{H}}.
\end{aligned}
\end{equation}
These 4 matrices are the central elements in subspace based SysID, where the past data denoted by
the subscript $p$ will be used to estimate the initial state, whereas the future data denoted by
the subscript $\text{f}$ will be used to predict the future trajectories. Note that all four matrices in Eq. (\ref{equpuf}) contain $\mathrm{N}$ row entries of the corresponding vectors, as we have chosen $\mathrm{M} = \mathrm{N}$. However, when $\mathrm{M} \neq \mathrm{N}$, the past matrices ($\mathbf{U}_{\text{p}}$ and $\mathbf{Y}_{\text{p}}$) will have $\mathrm{M}$ row entries, whereas the future matrices ($\mathbf{U}_{\text{f}}$ and $\mathbf{Y}_{\text{f}}$)  will have $\mathrm{N}$ row entries.
Using Eqs. (\ref{eqypred}), the elements of $\mathbf{Y}_{\mathrm{p}}$ and $\mathbf{Y}_{\mathrm{f}}$ can be computed as:
\begin{equation}
 \label{eqyd1}
    \begin{aligned}
        \mathbf{Y}_{1}&=\mathbf{A}_{\mathbf{Y}} \mathbf{x}_{0}+\mathbf{B}_{\mathbf{Y}} \mathbf{U}_{0}\\
        & \hspace{0.2cm} \vdots \\ \mathbf{Y}_{\mathrm{H}}&=\mathbf{A}_{\mathbf{Y}} \mathbf{x}_{{\mathrm{H}-1}}+\mathbf{B}_{\mathbf{Y}} \mathbf{U}_{{\mathrm{H}-1}}    \end{aligned}
\end{equation}
\begin{equation}
 \label{eqyd2}
    \begin{aligned}
        \mathbf{Y}_{\mathrm{N}+1}&=\mathbf{A}_{\mathbf{Y}} \mathbf{x}_{{\mathrm{N}}}+\mathbf{B}_{\mathbf{Y}} \mathbf{U}_{{\mathrm{N}}}\\& \hspace{0.2cm}\vdots\\
\mathbf{Y}_{{\mathrm{N}+\mathrm{H}}}&=\mathbf{A}_{\mathbf{Y}} \mathbf{x}_{{\mathrm{N}+\mathrm{H}-1}}+\mathbf{B}_{\mathbf{Y}} \mathbf{U}_{{\mathrm{N}+\mathrm{H}-1}}.
    \end{aligned}
\end{equation}
which results in:
\begin{equation}
\label{eqypyf}
    \begin{aligned}
        & \mathbf{Y}_{\text{p}}= \mathbf{A}_{\mathbf{Y}} \underbrace{\left[\begin{matrix} \mathbf{x}_{0} & \mathbf{x}_{1} & \dots & \mathbf{x}_{{\mathrm{H}-1}}\end{matrix}\right]}_{\mathbf{X}_{\text{p}}} + \mathbf{B}_{\mathbf{Y}} \mathbf{U}_{\text{p}}\\
        & \mathbf{Y}_{\text{f}}= \mathbf{A}_{\mathbf{Y}} \underbrace{\left[\begin{matrix} \mathbf{x}_{{\mathrm{N}}} & \mathbf{x}_{{\mathrm{N}+1}} & \dots & \mathbf{x}_{{\mathrm{N}+\mathrm{H}-1}}\end{matrix}\right]}_{\mathbf{X}_{\text{f}}} + \mathbf{B}_{\mathbf{Y}} \mathbf{U}_{\text{f}}
    \end{aligned}
\end{equation}
Using the solution of the state equation (last row of Eq.  (\ref{eqxpred})), the elements of $\mathbf{X}_{\mathrm{f}}$  can be computed as:
\begin{equation}
    \begin{aligned}
        \label{eqxdn}\mathbf{x}_{\mathrm{N}}&=\mathbf{A}^{\mathrm{N}} \mathbf{x}_{0}+\mathbf{A}^{\mathrm{N}-1}\mathbf{B} \mathbf{u}_{0}+\mathbf{A}^{\mathrm{N}-2}\mathbf{B} \mathbf{u}_{1}+\dots+ \mathbf{B} \mathbf{u}_{{\mathrm{N}-1}}=\mathbf{A}^{\mathrm{N}} \mathbf{x}_{0}+\mathbf{C}_{\mathbf{u}}\mathbf{U}_{0}\\
        & \hspace{0.2cm} \vdots \\         \mathbf{x}_{{\mathrm{N}+\mathrm{H}-1}}&=\mathbf{A}^{\mathrm{N}} \mathbf{x}_{{\mathrm{H}-1}}+\mathbf{A}^{\mathrm{N}-1}\mathbf{B} \mathbf{u}_{{\mathrm{H}-1}}+\mathbf{A}^{\mathrm{N}-2}\mathbf{B} \mathbf{u}_{{\mathrm{H}}}+\dots+ \mathbf{B} \mathbf{u}_{{\mathrm{N}+\mathrm{H}-2}}=\mathbf{A}^{\mathrm{N}} \mathbf{x}_{{\mathrm{H}-1}}+\mathbf{C}_{\mathbf{u}}\mathbf{U}_{{\mathrm{H}-1}}.
    \end{aligned}
\end{equation}
which can be compactly written as:
\begin{equation}
\label{eqxfxp}
    \mathbf{X}_{\mathrm{f}}= \mathbf{A}^{\mathrm{N}} \mathbf{X}_{\mathrm{p}} + \underbrace{\left[\begin{matrix} \mathbf{A}^{\mathrm{N}-1}\mathbf{B} & \mathbf{A}^{\mathrm{N}-2}\mathbf{B} & \dots & \mathbf{B}\end{matrix}\right]}_{\mathbf{C}_{\mathbf{u}}} \mathbf{U}_{\mathrm{p}}.
\end{equation} 
In \cite{bRP98}, it is shown that, for any observable system and $p\mathrm{N}\geq n,$ there exists a matrix $\mathbf{L}$ such that:
\begin{equation}
  \label{eqaz}    \mathbf{A}^{\mathrm{N}}+\mathbf{L} \mathbf{A}_{\mathbf{Y}}=0.
\end{equation}
Now, using Eqs. (\ref{eqypyf},\ref{eqxfxp},\ref{eqaz}), we can represent $\mathbf{Y}_{\mathrm{f}}$ as a linear function of $\mathbf{Y}_{\mathrm{p}}$, $\mathbf{U}_{\mathrm{p}}$ and $\mathbf{U}_{\mathrm{f}}$ which is the key idea in SPC.
This is achieved by substituting $\mathbf{X}_{\mathrm{f}}$ from Eq. (\ref{eqypyf}) in $\mathbf{Y}_{\mathrm{f}}$ and simplifying  results in:

\begin{equation}
\label{eqyzax}
\begin{aligned}
    \mathbf{Y}_{\mathrm{f}}&=\mathbf{A}_{\mathbf{Y}} \mathbf{A}^{\mathrm{N}} \mathbf{X}_{\mathrm{p}} + \mathbf{A}_{\mathbf{Y}}\mathbf{C}_{\mathbf{u}} \mathbf{U}_{\mathrm{p}} + \mathbf{B}_{\mathbf{Y}} \mathbf{U}_{\mathrm{f}}\\
    &= -\mathbf{A}_{\mathbf{Y}} \mathbf{L} \mathbf{A}_{\mathbf{Y}} \mathbf{X}_{\mathrm{p}} + \mathbf{A}_{\mathbf{Y}}\mathbf{C}_{\mathbf{u}} \mathbf{U}_{\mathrm{p}} + \mathbf{B}_{\mathbf{Y}} \mathbf{U}_{\mathrm{f}} \hspace{2.5cm} \mathbf{A}^{\mathrm{N}}= -\mathbf{L} \mathbf{A}_{\mathbf{Y}} \hspace{0.2cm}  \mathrm{from Eq. (\ref{eqaz})}  \\
    &=-\mathbf{A}_{\mathbf{Y}} \mathbf{L} [\mathbf{Y}_{\mathrm{p}}-\mathbf{B}_{\mathbf{Y}}\mathbf{U}_{\mathrm{p}}] + \mathbf{A}_{\mathbf{Y}}\mathbf{C}_{\mathbf{u}} \mathbf{U}_{\mathrm{p}} + \mathbf{B}_{\mathbf{Y}} \mathbf{U}_{\mathrm{f}}  \hspace{1.5cm} \mathbf{A}_{\mathbf{Y}} \mathbf{X}_{\mathrm{p}}= \mathbf{Y}_{\mathrm{p}}-\mathbf{B}_{\mathbf{Y}}\mathbf{U}_{\mathrm{p}}\hspace{0.2cm}  \mathrm{from Eq. (\ref{eqypyf})} \\
    &=\mathbf{A}_{\mathbf{Y}}[\mathbf{L}\mathbf{B}_{\mathbf{Y}}+\mathbf{C}_{\mathbf{u}}] \mathbf{U}_{\mathrm{p}} - \mathbf{A}_{\mathbf{X}}\mathbf{L} \mathbf{Y}_{\mathrm{p}} + \mathbf{B}_{\mathbf{Y}} \mathbf{U}_{\mathrm{f}}\\&=\mathbf{P}_{1}\mathbf{U}_{\mathrm{p}}+\mathbf{P}_{2}\mathbf{Y}_{\mathrm{p}}+\mathbf{B}_{\mathbf{Y}}\mathbf{U}_{\mathrm{f}}
    \end{aligned}
\end{equation}
where $\mathbf{P}_{1}=\mathbf{A}_{\mathbf{Y}}[\mathbf{L}\mathbf{B}_{\mathbf{Y}}+\mathbf{C}_{\mathbf{u}}]$, $\mathbf{P}_{2}=- \mathbf{A}_{\mathbf{Y}}\mathbf{L},$ $\mathbf{Y}_{\mathrm{f}} \in \mathbb{R}^{p\mathrm{N} \times \mathrm{H}},$ $\mathbf{P}_{1}\in \mathbb{R}^{p\mathrm{N} \times m\mathrm{N}}$, $\mathbf{U}_{\mathrm{p}}\in \mathbb{R}^{m\mathrm{N} \times \mathrm{H}}$, $\mathbf{P}_{2}\in \mathbb{R}^{p\mathrm{N} \times p\mathrm{N}}$, $\mathbf{Y}_{\mathrm{p}}\in \mathbb{R}^{p\mathrm{N} \times \mathrm{H}}$, $\mathbf{B}_{\mathbf{Y}}\in \mathbb{R}^{p\mathrm{N} \times m\mathrm{N}}$, $\mathbf{U}_{\mathrm{f}}\in \mathbb{R}^{m\mathrm{N} \times \mathrm{H}}$. 
By defining
\begin{equation}
\label{eqwp}
    \mathbf{P}=\left[\begin{matrix} \mathbf{P}_{1} & \mathbf{P}_{2} &  \mathbf{B}_{\mathbf{Y}}\end{matrix}\right] \in \mathbb{R}^{p\mathrm{N} \times (m\mathrm{N}+p\mathrm{N} + m\mathrm{N}) }  \hspace{1cm}\mathbf{S}= \left[\begin{matrix} \mathbf{U}_{\mathrm{p}} \\ \mathbf{Y}_{\mathrm{p}} \\  \mathbf{U}_{\mathrm{f}}\end{matrix}\right] \in \mathbb{R}^{ (m\mathrm{N}+p\mathrm{N} + m\mathrm{N})  \times \mathrm{H} } 
\end{equation}
  Eq. (\ref{eqyzax}) can be compactly written as:
\begin{equation}
    \label{eqyzp} \mathbf{Y}_{\mathrm{f}}=\mathbf{P}\mathbf{S}
\end{equation}
which gives the parameter matrix $\mathbf{P}$ as:
\begin{equation}
    \label{eqpy} \mathbf{P}=\mathbf{Y}_{\mathrm{f}}\mathbf{S}^{\dagger}
\end{equation}
where $\mathbf{S}^{\dagger}=(\mathbf{S}^{\top}\mathbf{S})^{-1}\mathbf{S}^{\top}$ is the pseudoinverse of $\mathbf{S}$. Note that Eq. (\ref{eqyzp}) shows that the future output can be represented as a linear function of the past outputs and past inputs. The identified parameters $\mathbf{P}_{1},\mathbf{P}_{2},\mathbf{B}_{\mathbf{Y}}$ can be used for computing the predicted output sequence $\mathbf{Y}_{k+1}$ in Eq. (\ref{eqypred}) by defining 
\begin{equation} 
\label{eqypkupk}
\begin{aligned}
&\mathbf{Y}_{k+1}=\mathbf{Y}_{\mathrm{f}_1}=\mathbf{H}_{(\mathrm{N},1,\mathbf{y}_{k+1|k})}   \hspace{0.8cm} \mathbf{Y}_{\mathrm{p}_1}=\mathbf{H}_{(\mathrm{N},1,\mathbf{y}_{k-\mathrm{N}+1|k})}\\
&\mathbf{U}_{k}=\mathbf{U}_{\mathrm{f}_1}=\mathbf{H}_{(\mathrm{N},1,\mathbf{u}_{k|k})} \hspace{1.45cm}  \mathbf{U}_{\mathrm{p}_1}=\mathbf{H}_{(\mathrm{N},1,\mathbf{u}_{k-\mathrm{N}+1|k})}
\end{aligned}
\end{equation}
Now, using Eq. (\ref{eqyzax}), $\mathbf{Y}_{k+1}$ can be represented as:
\begin{equation}
  \label{eqspcconstr}  \mathbf{Y}_{k+1}=\mathbf{P}_{1}\mathbf{U}_{\mathrm{p}_1}+\mathbf{P}_{2}\mathbf{Y}_{\mathrm{p}_1}+\mathbf{B}_{\mathbf{Y}}\mathbf{U}_{k}
\end{equation}
This equality constraint  can be used for replacing the model-based constraint in Eq. (\ref{eqlmpc}).
By comparing Eq. (\ref{eqspcconstr}) with Eq. (\ref{eqypred}), it can be observed that the terms $\mathbf{P}_{1}\mathbf{U}_{\mathrm{p}_1}+\mathbf{P}_{2}\mathbf{Y}_{\mathrm{p}_1}$ and 
$\mathbf{A}_{\mathbf{Y}} \mathbf{x}_{k|k}$ are equivalent. This equivalence is further supported by the definition of observability, which states that the initial state can be estimated from a sufficiently long sequence of input and output measurements. 
 Using the predicted output sequence in Eq. (\ref{eqspcconstr}), the optimization problem for SPC can be defined as:
\begin{equation}
\label{eqoptspc}
\begin{aligned}
 \underset{\mathbf{U}_{k}}{\inf}  ~&~  [\mathbf{Y}_{r}-(\mathbf{P}_{1}\mathbf{U}_{\mathrm{p}_1}+\mathbf{P}_{2}\mathbf{Y}_{\mathrm{p}_1}+\mathbf{B}_{\mathbf{Y}}\mathbf{U}_{k})]^{\top}\mathbf{Q}_{\mathbf{Y}} [\mathbf{Y}_{r}-(\mathbf{P}_{1}\mathbf{U}_{\mathrm{p}_1}+\mathbf{P}_{2}\mathbf{Y}_{\mathrm{p}_1}+\mathbf{B}_{\mathbf{Y}}\mathbf{U}_{k})] + \mathbf{U}_{k}^{\top} \mathbf{R}_{\mathbf{U}}\mathbf{U}_{k} \\
\mathrm{subject \hspace{0.1cm}to} ~&~  \mathbf{P}_{1}\mathbf{U}_{\mathrm{p}_1}+\mathbf{P}_{2}\mathbf{Y}_{\mathrm{p}_1}+\mathbf{B}_{\mathbf{Y}}\mathbf{U}_{k} \in \mathbb{Y}^{\mathrm{N}}, \hspace{0.2cm}
\mathbf{U}_{k} \in \mathbb{U}^{\mathrm{N}}.
     \end{aligned}
\end{equation}
Note that in SPC, the past data is used by the SysID algorithm for identifying a model as in Eq. (\ref{eqyzax}), and the future data predicted by the model is used in MPC for optimizing the system behavior, i.e., SPC is a combination of subspace-based SysID and MPC. Eventhough we considered the past horizon same as future horizon $\mathrm{M}=\mathrm{N}$ in the derivation, the same model as in Eq. (\ref{eqyzax}) can be used for $\mathrm{M}\neq \mathrm{N}$, where $\mathbf{A}_{\mathbf{Y}}$ for $\mathbf{U}_{\mathrm{p}}$ and $\mathbf{Y}_{\mathrm{p}}$ in Eq. (\ref{eqypyf})  is to be replaced with $\mathbf{A}_{\mathbf{Y}_{\mathrm{M}}}$.  The algorithm for SPC is summarized below.
   
    \begin{algorithm}[H]
 
	\begin{algorithmic}[1] 
	
	\STATE Require the input-output data $\mathbf{Y}_{\mathrm{D}} $ and $\mathbf{U}_{\mathrm{D}}$.
    \STATE Select $\mathrm{N},$ $\mathrm{M}$, and $\mathrm{H}$ such that $\mathrm{N}+\mathrm{M}+\mathrm{H}-1 \leq \mathrm{D}$. 
		\STATE Compute $\mathbf{U}_{\mathrm{p}},$ $\mathbf{Y}_{\mathrm{p}}$ $\mathbf{U}_{\mathrm{f}}$ and $\mathbf{Y}_{\mathrm{f}}$ as in Eq. (\ref{equpuf}).
		\STATE Compute $\mathbf{P}$ (or $\mathbf{P}_{1},\mathbf{P}_{2},\mathbf{B}_{\mathbf{Y}}$) as in Eq. (\ref{eqpy})     
         \STATE Select $\mathrm{N}_{\mathrm{T}}$ $\mathbf{y}_{\mathrm{r}},$  $\mathbf{Q}_{\mathbf{y}}$, $\mathbf{R}_{\mathbf{u}}.$ Construct $\mathbf{Q}_{\mathbf{Y}}$ and $\mathbf{R}_{\mathbf{U}}$ as in Eq. (\ref{eqqyru})
    \STATE Initialize $\mathbf{U}_{k}$
	\FOR  {$k= 0~to~ \mathrm{N}_{\mathrm{T}}-1 $}
        \STATE Compute $\mathbf{Y}_{r}$ as in Eq. (\ref{eqqyru})
		\STATE Compute $\mathbf{Y}_{\mathrm{p}_1}$ and $\mathbf{U}_{\mathrm{p}_1}$ as in Eq. (\ref{eqypkupk})
		\STATE  Compute $\mathbf{U}_{k}^{*}$ by solving the optimization problem in Eq. (\ref{eqoptspc})
		\STATE Apply  $\mathbf{u}_{k|k}=\mathbf{U}_{k_{(1:m)}}^{*}$ to the system 
		\ENDFOR
    
	\end{algorithmic}
	\caption{: SPC}
    \label{algspc}
\end{algorithm}

\subsubsection{SPC: Numerical example}
\label{secspc}
\begin{figure}[H] 
 		\begin{center}
 		\includegraphics [scale=.165] {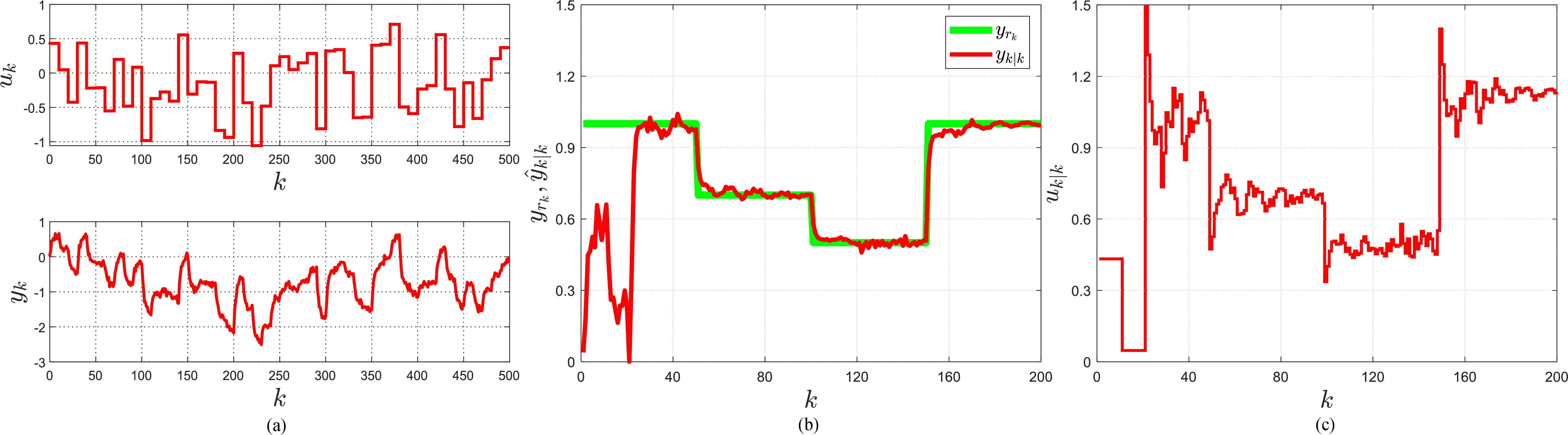}
 		\caption{{ SPC \hspace{.15cm} (a) Training input and output \hspace{.1cm} (b) Output response with SPC\hspace{.1cm}(c) Control input.}}
        \label{figspc}
 	\end{center}
    
 \end{figure}

To illustrate SPC, the input-output training data is generated using a fourth-order LTI system for which the parameters are chosen as follows:
\begin{equation}
\label{eqspcsim}
    \mathbf{A}=\left[\begin{matrix} 0.5 &0 & 0.05 & 0.1 \\ 0 & 0.7 & 0.6 & 0.4 \\ 0.1 & 0.2 & 0.5 & 0.1 \\ 0.2 & 0.1 & -0.1 & 0.1
\end{matrix}\right] \hspace{1cm} \mathbf{B}=\left[\begin{matrix} 0.5 \\0.2 \\ 0.1 \\ 0.7
\end{matrix}\right]  \hspace{1cm} \mathbf{C}=\left[\begin{matrix} 1 & 0 & 0 & 0
\end{matrix}\right] .
\end{equation}
The LTI system is simulated for $1000$ instants with a multilevel pseudorandom sequence (PRMS) input, and the first $500$ samples are used for training the model, i.e., $\mathrm{D}=500$. Here, PRMS input with a sufficient number of perturbations is used so that the input is persistently exciting, i.e., sufficiently rich to capture system dynamics. See \cite{bJW05} for more details of the persistency of excitation.   Using the training data, the model-based constraint parameters: $\mathbf{P}_{1}, \mathbf{P}_{2},$ and $\mathbf{B}_{\mathbf{Y}}$ are computed with $\mathrm{N}=30,$ $\mathrm{M}=20$ and $\mathrm{H}=400$ which are then used in MPC as in Algorithm \ref{algspc}.   
 The simulation parameters are chosen as $\mathrm{N}_{\mathrm{T}}=200,$ $\mathrm{N}=30,$ $\mathbf{Q}_{\mathbf{y}}=5,\mathbf{R}_{\mathbf{u}}=0.1$ and $\mathbf{x}_{0}=\left[\begin{matrix} 0 & 0 & 0 & 0
\end{matrix}\right]^{T}.$ The constraint set is defined as $u_{min}=-5,$ $u_{max}=5$.
The output reference is chosen as $\mathbf{y}_{\mathrm{r}}=1$ for the first quadrant (first 50 instants), which is then changed into $0.7,$ $0.5,$ and back to $1$ in each successive quadrant.  The training input and output data used for identifying the model parameters are plotted in Fig. \ref{figspc}(a). The simulation response for the output tracking is given in Fig. \ref{figspc}(b), which shows that the output follows the desired reference.
    During MPC implementation, the first $\mathrm{M}$ instants (past data) of the output are chosen from the training output, and the MPC problem is solved from $k=\mathrm{M}+1$ onward. That's why the output for the initial 30 instants in Fig. \ref{figspc} does not follow the reference $\mathbf{y}_{\mathrm{r}_k}$. The control input plot is given in Fig. \ref{figspc}(c). 

\subsection{Data-enabled Predictive Control (DeePC)}
\label{secdeepc}
Data-enabled predictive control (DeePC) is a model-free control approach builds upon the foundational concepts introduced in SPC.  This approach utilizes the Hankel matrix to organize historical input-output trajectories, allowing the prediction of future outputs through linear combinations of past measured data. The DeePC uses the same Hankel matrices $\mathbf{U}_{\mathrm{p}}$, $\mathbf{U}_{\mathrm{f}}$, $\mathbf{Y}_{\mathrm{p}}$, and $\mathbf{Y}_{\mathrm{f}}$ as in Eq. (\ref{equpuf}).  Similar to SPC, we define $\mathbf{Y}_{k+1}=\mathbf{Y}_{\mathrm{f}_1}$, $\mathbf{U}_{k}=\mathbf{U}_{\mathrm{f}_1}$, $\mathbf{Y}_{\mathrm{p}_1}$, and $\mathbf{U}_{\mathrm{p}_1}$ as in Eq. (\ref{eqspcconstr}).
SPC was associated with a model identification step where we used the Hankel matrix equation (\ref{eqyzax}) to identify the parameters $\mathbf{P}_{1}$, $\mathbf{P}_{2}$, and $\mathbf{B}_{\mathbf{Y}}$ which are then used to define the equality constraint in the optimization problem for SPC. In contrast, the DeePC eliminates the model identification step altogether. Instead of relying on a parametric model, DeePC directly utilizes measured input-output data in Hankel matrix form to construct the equality constraints required for the predictive control problem. This is made possible through the application of the \textbf{fundamental lemma} \cite{bJW05} from behavioral systems theory, which provides a data-driven mechanism to represent all system trajectories using persistently exciting data. By doing so, DeePC enables a fully non-parametric and model-free formulation of predictive control, which retains the performance and structure of traditional model-based approaches while enhancing adaptability and ease of implementation. The fundamental lemma, which forms the theoretical backbone of this data-driven approach, is presented next. 

\begin{definition}[\cite{bJW05}]
    The input sequence $\{\mathbf{u}_{0},\mathbf{u}_{1},\dots,\mathbf{u}_{\mathrm{D}-1}\}$ is persistently exciting (PE) with order $\mathrm{L}$ if $\mathbf{H}_{(\mathrm{L},\mathrm{H},\mathbf{u}_{0})}$ has full row rank, i.e., rank of $\mathbf{H}_{(\mathrm{L},\mathrm{H},\mathbf{u}_{0})}$ $=$ $m\mathrm{L}.$
\end{definition}
\begin{lemma}[Fundamental lemma \cite{bJW05}]
    Suppose $\{\mathbf{u}_{0},\mathbf{u}_{1},\dots,\mathbf{u}_{\mathrm{D}-1}\}$ is persistently exciting of order $\mathrm{N}+\mathrm{M}+n$. Then for every  trajectory $(\mathbf{Y}_{k+1},\mathbf{U}_{k})$ of the system there exists a solution $\mathbf{v}_{k} \in \mathbb{R}^{\mathrm{H}}$ satisfies: 
    \begin{equation}
    \label{eqfund_lemma}
     \left[\begin{matrix} \mathbf{U}_{\mathrm{p}} \\ \mathbf{Y}_{\mathrm{p}} \\ \mathbf{U}_{\mathrm{f}} \\ \mathbf{Y}_{\mathrm{f}}
\end{matrix}\right]\mathbf{v}_{k} = \left[\begin{matrix} \mathbf{U}_{\mathrm{p}_1} \\ \mathbf{Y}_{\mathrm{p}_1} \\ \mathbf{U}_{k} \\ \mathbf{Y}_{k+1}
\end{matrix}\right].
    \end{equation}
\end{lemma}
The Eq. (\ref{eqfund_lemma}) given by fundamental lemma can be considered as data-driven equality constraint which can be used instead of model-based equality constraints. 
The state space model in the MPC optimization problem in Eq. (\ref{eqlmpc}) or the model-based equality constraint in Eq. (\ref{eqoptympc2}) can be replaced by a data-dependent linear equality constraint defined with Hankel
 matrices as in Eq. (\ref{eqfund_lemma}). By doing so, DeePC effectively bypasses the need for an intermediate system identification step, offering a powerful and flexible framework for predictive control in both linear and certain nonlinear systems. In this context, DeePC can be regarded as a promising method within the broader scope of adaptive control.  The 
optimization problem for DeePC is defined as:
\begin{equation}
\label{eqoptdpc}
\begin{aligned}
 \underset{\mathbf{U}_{k},\mathbf{Y}_{k+1},\mathbf{v}_{k}}{\inf}  ~&~ [\mathbf{Y}_{r}-\mathbf{Y}_{k+1}]^{\top}\mathbf{Q}_{\mathbf{Y}} [\mathbf{Y}_{r}-\mathbf{Y}_{k+1}] + \mathbf{U}_{k}^{\top} \mathbf{R}_{\mathbf{U}}\mathbf{U}_{k}+\alpha L(\mathbf{v}_{k}) \\
\mathrm{subject \hspace{0.1cm}to} ~&~   \mathbf{Y}_{k+1}\in \mathbb{Y}^{\mathrm{N}}  ,\mathbf{U}_{k} \in \mathbb{U}^{\mathrm{N}}   \\
  ~&~  \left[\begin{matrix} \mathbf{U}_{\mathrm{p}} \\ \mathbf{Y}_{\mathrm{p}} \\ \mathbf{U}_{\mathrm{f}} \\ \mathbf{Y}_{\mathrm{f}}
\end{matrix}\right]\mathbf{v}_{k} = \left[\begin{matrix} \mathbf{U}_{\mathrm{p}_1} \\ \mathbf{Y}_{\mathrm{p}_1} \\ \mathbf{U}_{k} \\ \mathbf{Y}_{k+1}
\end{matrix}\right] 
     \end{aligned}
\end{equation}
where $L(\mathbf{v}_{k})$ is a regularization term included in the cost function for DeePC to avoid undesirable solutions which is commonly chosen as  $L(\mathbf{v}_{k})={\parallel \mathbf{v}_{k} \parallel}_{2}^{2}$. It is important to note that, in the optimization problem for DeePC, the predicted output sequence $\mathbf{Y}_{k+1}$ is also treated as a decision variable. This is because, in the absence of an explicit system model, $\mathbf{Y}_{k+1}$ cannot be expressed directly as a function of the control input sequence $\mathbf{U}_{k}$ and therefore cannot be eliminated from the cost function. This increases the number of decision variables in the optimization problem, which affects its application to high dimension systems. Additionally, the lack of an explicit system model complicates the handling of nonlinearities, ensuring theoretical guarantees such as robustness, stability, optimality, etc. 
 The algorithm for DeePC is summarized below.
   
    \begin{algorithm}[H]
 
	\begin{algorithmic}[1] 
	
	\STATE Require the input-output data $\mathbf{Y}_{\mathrm{D}} $ and $\mathbf{U}_{\mathrm{D}}$.
    \STATE Select $\mathrm{N},$ $\mathrm{M}$, and $\mathrm{H}$ such that $\mathrm{N}+\mathrm{M}+\mathrm{H}-1 \leq \mathrm{D}$. 
		\STATE Compute $\mathbf{U}_{\mathrm{p}},$ $\mathbf{Y}_{\mathrm{p}}$ $\mathbf{U}_{\mathrm{f}}$ and $\mathbf{Y}_{\mathrm{f}}$ as in Eq. (\ref{equpuf}).   
         \STATE Select $\alpha,$ $\mathrm{N}_{\mathrm{T}}$ $\mathbf{y}_{\mathrm{r}},$  $\mathbf{Q}_{\mathbf{y}}$, $\mathbf{R}_{\mathbf{u}}.$ Construct $\mathbf{Q}_{\mathbf{Y}}$ and $\mathbf{R}_{\mathbf{U}}$ as in Eq. (\ref{eqqyru})
    \STATE Initialize $\mathbf{U}_{k}$
	\FOR  {$k= 0~to~ \mathrm{N}_{\mathrm{T}}-1 $}
        \STATE Compute $\mathbf{Y}_{r}$ as in Eq. (\ref{eqqyru})
		\STATE Compute $\mathbf{Y}_{\mathrm{p}_1}$ and $\mathbf{U}_{\mathrm{p}_1}$ as in Eq. (\ref{eqypkupk})
		\STATE  Compute $\mathbf{U}_{k}^{*}$ by solving the optimization problem in Eq. (\ref{eqoptdpc})
		\STATE Apply  $\mathbf{u}_{k|k}=\mathbf{U}_{k_{(1:m)}}^{*}$ to the system 
		\ENDFOR
    
	\end{algorithmic}
	\caption{: DeePC}
    \label{algdeepc}
\end{algorithm}

\subsubsection{DeePC: Numerical example}
\begin{figure}[H] 
   		\begin{center}
 		\includegraphics [scale=.165] {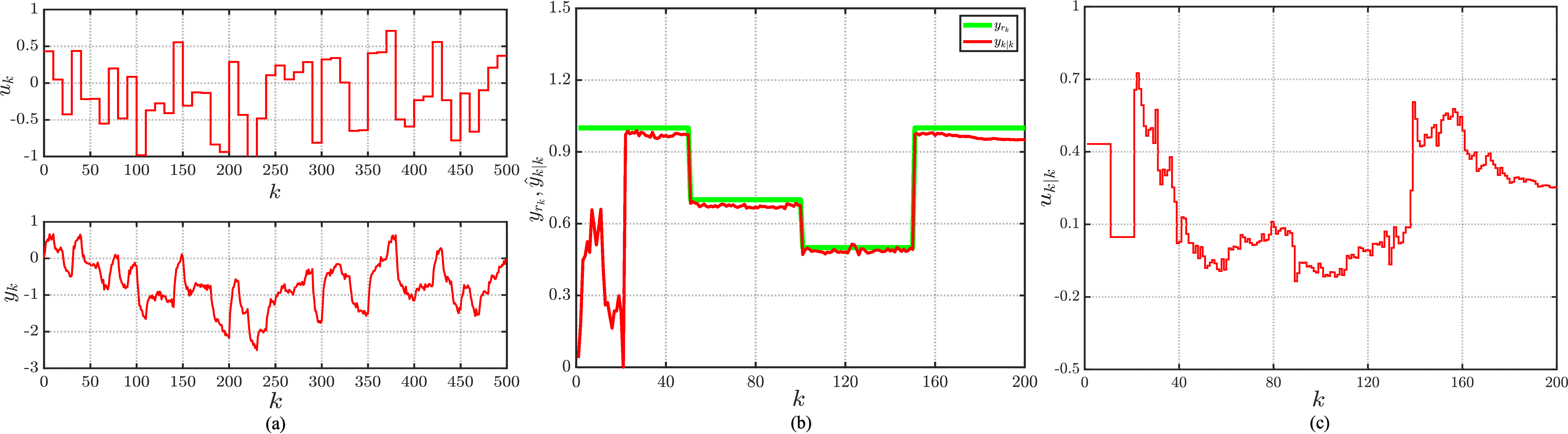}
 		\caption{ DeePC \hspace{.15cm} (a) Training input and output \hspace{.1cm} (b) Output response with DeePC\hspace{.1cm}(c) Control input.}
        \label{figdeepc}
  	\end{center}
   \end{figure}

To demonstrate the DeePC approach, the same example presented for SPC in Section \ref{secspc} is considered, with the system parameters defined as in Equation (\ref{eqspcsim}). The training data used to construct the Hankel matrices is identical, with the parameters set to $\mathrm{N} = 30$, $\mathrm{M} = 20$, and $\mathrm{H} = 200$. Based on the resulting input-output Hankel matrices, the DeePC algorithm is implemented as outlined in  Algorithm \ref{algdeepc}. 
 The simulation parameters are selected as follows: $\mathrm{N}_{\mathrm{T}} = 200$, $\mathrm{N} = 30$, $\mathbf{Q}_{\mathbf{y}} = 5$, $\mathbf{R}_{\mathbf{u}} = 0.1$, and $\alpha =1$. The input constraints are defined by $u_{\min} = -5$ and $u_{\max} = 5$. Fig. \ref{figdeepc}(a) illustrates the input-output training data used for constructing the Hankel matrices. The simulation results for output tracking using DeePC are presented in Fig. \ref{figdeepc}(b), demonstrating that the system output effectively tracks the desired reference signal. The corresponding control input is shown in Fig. \ref{figdeepc}(c).
\subsection{PEM based D-LMPC}
 Prediction error methods (PEMs) are optimization-based SysID approaches in which the model parameters are estimated by minimizing a function of the prediction error \cite{bRD17}.   
 The training data is considered to be available in the form of input-output samples as in Eq. (\ref{equy}). Let the unknown parameters to be identified:   $\mathbf{A},$ $\mathbf{B},$ $\mathbf{C}$ and $\mathbf{x}_{0}$ be stored in a parameter matrix as:
\begin{equation}
    \bm{\theta}_{\mathbf{x}}= \left[\begin{matrix} \mathbf{A} & \mathbf{B} & \mathbf{C}^{\top} & \mathbf{x}_{0} \end{matrix}\right].
\end{equation}
The predicted state and  output sequence for the training data can be computed as:
\begin{equation}
 \label{eqxnny}
\begin{aligned}
&\mathbf{X}_{\mathrm{D}}=\left[\begin{matrix} \mathbf{A}\mathbf{x}_{0}+\mathbf{B}\mathbf{u}_{0}& \mathbf{A}^{2}\mathbf{x}_{0} +\mathbf{A}\mathbf{B}\mathbf{u}_{0} + \mathbf{B}\mathbf{u}_{1} & \dots &  \mathbf{A}^{\mathrm{D}}\mathbf{x}_{0} +\mathbf{A}^{\mathrm{D}-1}\mathbf{B}\mathbf{u}_{0} + \dots + \mathbf{B}\mathbf{u}_{{\mathrm{D}-1}}   \end{matrix}\right]\\
&\hat{\mathbf{Y}}_{\mathrm{D}}=\mathbf{C}\mathbf{X}_{\mathrm{D}}= \mathbf{f}_{\mathrm{yp}}(\bm{\theta}_{\mathbf{x}}).
\end{aligned}
\end{equation}
The loss function for PEM is defined as
\begin{equation}
\label{eqjy}
L_{y}= \sum_{k=1}^{\mathrm{D}}  {\parallel \mathbf{y}_{k} - \hat{\mathbf{y}}_{k} \parallel }^{2}=  {\parallel \mathbf{Y}_{\mathrm{D}}-\hat{\mathbf{Y}}_{\mathrm{D}} \parallel }_{F}^{2}=\operatorname{trace}([\mathbf{Y}_{\mathrm{D}}-\mathbf{f}_{\mathrm{yp}}(\bm{\theta}_{\mathbf{x}})]^{\top}[\mathbf{Y}_{\mathrm{D}}-\mathbf{f}_{\mathrm{yp}}(\bm{\theta}_{\mathbf{x}})]).
\end{equation}
The optimization problem for PEM-based linear state-space model identification becomes:
\begin{equation}
\label{eqpem}
\underset{\bm{\theta}_{\mathbf{x}}}{\min} \hspace{.2cm}  L_{y}.
\end{equation} 
Note that the SysID part can be performed either offline (only once before implementing the MPC scheme) or online (at each time instant or whenever required based on some event-based criteria).
The optimal model parameter: $\bm{\theta}_{\mathbf{x}}^{*}$  
 identified by solving the above optimization problem can be used for designing output or state-based LMPC as in Section \ref{seclmpc} which results in PEM-based D-LMPC. Note that since state-space models are not unique, the system parameters that fit a given input-output data are not unique and depends on the initial state. The optimal parameters $\mathbf{A}^{*},\mathbf{B}^{*},\mathbf{C}^{*}$ obtained by PEM is corresponding to a specific initial condition $\mathbf{x}_{0}^{*}$ and therefore may not match with actual system parameters $\mathbf{A},\mathbf{B},\mathbf{C}$.    Here, we are considering a state-based LMPC implementation for which the state and control references $\mathbf{x}_{\mathrm{r}},$ $\mathbf{u}_{\mathrm{r}}$ are computed using the steady-state criteria: $\mathbf{x}_{k+1}=\mathbf{x}_{k}=\mathbf{x}_{\mathrm{r}}$ and $\mathbf{y}_{k}=\mathbf{y}_{\mathrm{r}}$ resulting in:
\begin{equation}
\label{eqlmpcref}
\left[\begin{matrix}  \mathbf{x}_{\mathrm{r}}-\mathbf{A}^{*}\mathbf{x}_{\mathrm{r}}+ \mathbf{B}^{*}\mathbf{u}_{\mathrm{r}}  \\
\mathbf{y}_{\mathrm{r}}-\mathbf{C}^{*}\mathbf{x}_{\mathrm{r}}   \end{matrix}\right]
=\mathbf{0}.
    \end{equation}
Solving the above equation for a known $\mathbf{y}_{r}$ gives $\mathbf{x}_{\mathrm{r}}$ and $\mathbf{u}_{\mathrm{r}}$ which can be used to implement state-based LMPC. The algorithm for PEM-based D-LMPC is given below:
\begin{algorithm}[H]
 
	\begin{algorithmic}[1] 
	
	\STATE Require $\mathbf{U}_{\mathrm{D}}$ and $\mathbf{Y}_{\mathrm{D}}$
     \STATE Select $\mathrm{N}$
     \STATE Initialize $\bm{\theta}_{\mathbf{x}}$
\STATE Compute $\bm{\theta}_{\mathbf{x}}^{*}$  
 (contains $\mathbf{A}^{*},\mathbf{B}^{*},\mathbf{C}^{*}$) by solving Eq. (\ref{eqpem}).  
   \STATE Select $\mathrm{N},$ $\mathbf{y}_{\mathrm{r}},$  $\mathbf{Q}_{\mathbf{x}}$, $\mathbf{R}_{\mathbf{u}}.$ Construct $\mathbf{Q}_{\mathbf{X}}$ and $\mathbf{R}_{\mathbf{U}}$ as in Eq.  (\ref{eqqyru}) 
     \STATE Compute $\mathbf{x}_{\mathrm{r}}$ and $\mathbf{u}_{\mathrm{r}}$ by solving Eq. (\ref{eqlmpcref})
    \STATE Initialize $\mathbf{U}_{k}$
	\FOR  {$k= 0~to~ \mathrm{N}_{\mathrm{T}}-1 $}
        \STATE Compute $\mathbf{X}_{r}$ and $\mathbf{U}_{\mathrm{r}}$  as in Eq. (\ref{eqqyru}) 
		\STATE  Compute $\mathbf{U}_{k}^{*}$ by solving Eq. (\ref{eqnmpc2y}) with $\mathbf{x}_{k|k}$ is replaced by $\mathbf{y}_{k|k}$
		\STATE Apply  $\mathbf{u}_{k|k}=\mathbf{U}_{k_{(1:m)}}^{*}$ to the system  
		\ENDFOR
    
	\end{algorithmic}
	\caption{: PEM-based D-LMPC}
\end{algorithm}
\subsubsection{PEM-based D-LMPC: Numerical example}
\begin{figure}[H] 
 		\begin{center}
 		\includegraphics [scale=.165] {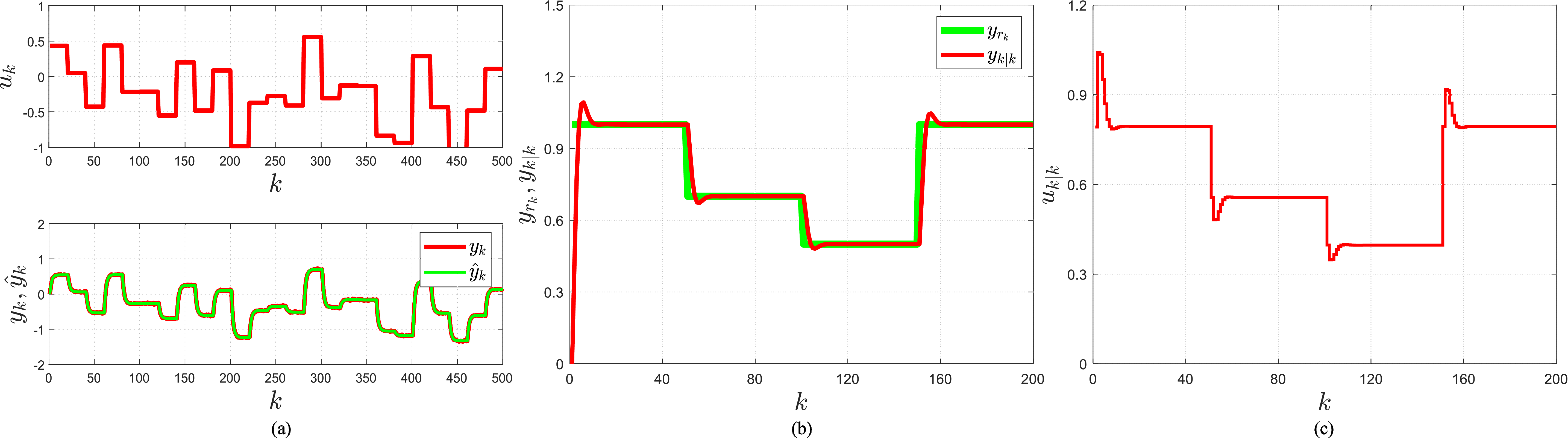}
 		\caption{{ PEM-based D-LMPC \hspace{.15cm} (a) Training input and output  \hspace{.1cm}(b) Output with D-LMPC \hspace{.1cm}(b) Control input.}}	
            \label{figpem}
 	\end{center}

 \end{figure}

To illustrate PEM-based D-LMPC, the input-output training data is generated using a fourth-order LTI system for which the parameters are:
\begin{equation}
    \mathbf{A}=\left[\begin{matrix} 0.5 &0 & 0.05 & 0.1 \\ 0 & 0.7 & 0 & 0.04 \\ 0 & 0 & 0.55 & 0.1 &\\ 0.2 & 0.1 & 0 & 0.1
\end{matrix}\right] \hspace{1cm} \mathbf{B}=\left[\begin{matrix} 0.5 \\0 \\ 0.1 \\ 0.7
\end{matrix}\right]  \hspace{1cm} \mathbf{C}=\left[\begin{matrix} 1 & 0 & 0 & 0
\end{matrix}\right] .
\end{equation}
The LTI system is simulated for $1000$ instants with PRMS input, and the first $500$ samples are used for training the model and the rest for validating the model. The system parameters identified with PEM is then used for implementing MPC for which the simulation parameters are chosen as $\mathrm{N}_{\mathrm{T}}=200,$ $\mathrm{N}=20,$ $\mathbf{Q}_{\mathbf{x}}=2\mathbf{I}_{4},\mathbf{R}_{\mathbf{u}}=3$ and $\mathbf{x}_{0}=\left[\begin{matrix} 0 & 0 & 0 & 0
\end{matrix}\right]^{T}.$ The constraint set is defined as $u_{min}=-5,$ $u_{max}=5$.
    The response of the LTI system with the PEM-based D-LMPC scheme is given in Fig. \ref{figpem}. The output reference is chosen similarly to the SPC example, with $\mathbf{y}_{\mathrm{r}}$ taking values as $1,0.7,0.5$.  The training input and output data, along with the output predicted with the PEM model, are plotted in Fig. \ref{figpem}(a). The output tracking with MPC is given in Fig. \ref{figpem}(b), which shows that the output follows the desired reference.
     The control input plot is given in Fig. \ref{figpem}(c). 

With this, we conclude our discussion on D-LMPC approaches. In the following section, we shift our focus to nonlinear SysID approaches and D-NMPC.
\section{Data-driven Nonlinear MPC}
\begin{figure}[H]
 		\begin{center}
 		\includegraphics [scale=.6] {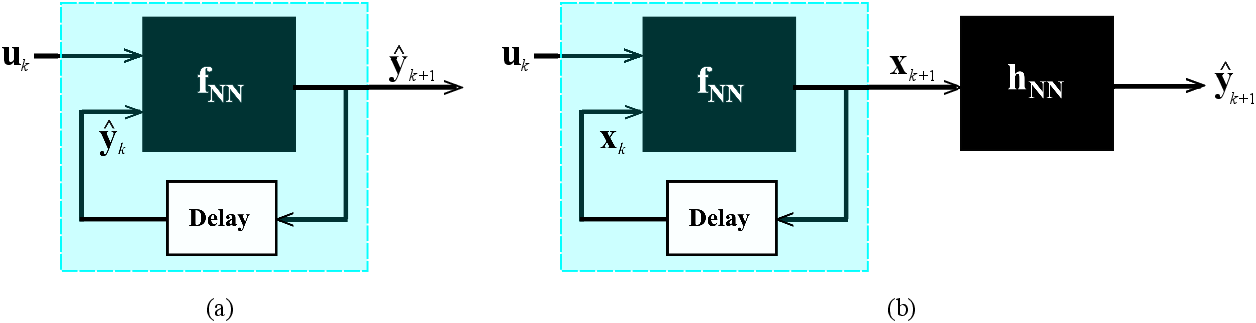}
 		\caption{Block diagram: (a) RNN \hspace{0.1cm} (b) SSNN  \cite{bMMSTF24}.}
 		\label{figSSNN}		
 	\end{center}
 \end{figure}

Data-driven NMPC (D-NMPC) is used for controlling systems that are nonlinear and have a wide operating range so that a linear model is not sufficient for accurately predicting the output. In D-NMPC, the system and output functions  $\mathbf{f}$ and $\mathbf{h}$ in Eq. (\ref{eqnlfx}) are considered to be unknown; instead, an input-output sequence $\mathbf{U}_{\mathrm{D}},\mathbf{Y}_{\mathrm{D}}$ as in Eq. (\ref{equy}) is available.  Most of the existing model-based D-NMPC schemes are using PEM, i.e., PEM-based SysID followed by NMPC. In the rest of the paper, we discuss PEM-based D-NMPC and its numerical implementation, where we focus on neural network (NN) based models:
\begin{enumerate}
    \item \textbf{Recurrent neural network} (RNN): is a neural network-based input-output dynamic model. The recurrent (or feedback) connections with delay make RNN a dynamic model and can be represented as a difference equation. (See Fig. \ref{figSSNN}(a)).
    \item \textbf{State-space neural network} (SSNN): is a data-driven state space model in which the state and output functions  are approximated using NNs (See Fig. \ref{figSSNN}(b)). SSNN comes under the class of RNNs.
\end{enumerate}

\subsection{Recurrent Neural Network based D-NMPC}
Recurrent neural networks (RNNs) are input-output dynamic models used to represent nonlinear systems \cite{bKK90,bRD95}. RNNs are based on NNs with a recurrent or feedback connection, which results in dynamic prediction. Given a training data in the form of input-output sequence as in Eq. (\ref{equy}), RNNs compute the predicted output using a recurrent equation:
\begin{equation} 
\label{eqyhatk1}
\hat{\mathbf{y}}_{k+1}=\mathbf{f}_{\mathrm{NN}}(\hat{\mathbf{y}}_{k},\mathbf{u}_{k};\bm{\theta}_{\mathbf{f}})=
\mathbf{f}_{\mathrm{NN}}(\hat{\mathbf{y}}_{k},\mathbf{u}_{k})
\end{equation}
where $\bm{\theta}_{\mathbf{f}} \in \mathbb{R}^{n_f} $  contains the training parameters of RNN, which consists of the weights and biases of the network. The block diagram of RNN is given in Fig. \ref{figSSNN}(a) in which the \textbf{delay} term is equivalent to the derivative in continuous time, which makes RNN a dynamic model. This represents one of the simplest forms of RNNs and can be viewed as being inspired by the initial versions of RNNs such as the Hopfield network \cite{bJH84}.  In general, RNN models can be represented as a composite function:
\begin{equation}
 \label{eqfnngnn}
\mathbf{f}_{\mathrm{NN}}=\mathbf{f}_{\mathrm{L}}(\mathbf{f}_{\mathrm{L}-1}(\dots(\mathbf{f}_{1}(\hat{\mathbf{y}}_{k},\mathbf{u}_{k}))))
\end{equation}
 where $\mathbf{f}_{i}, i=1,2,\dots,\mathrm{L}$ are layers of the RNN which can be represented as:
\begin{equation}
\label{eqfigi}
    \mathbf{f}_{i}(\mathbf{v}_{i})=\bm{\sigma}_{\mathbf{f}_i}(\mathbf{A}_{\mathbf{f}_i}\mathbf{v}_{i}+\mathbf{b}_{\mathbf{f}_i})
\end{equation}
where 
$\bm{\sigma}_{\mathbf{f}_i}:\mathbb{R}^{n_i}\rightarrow \mathbb{R}^{n_{i}}$ contains the activation functions for the $n_{i}$ neurons in the $i^{th}$ layer of the RNN. The corresponding weight and bias parameters being $\mathbf{A}_{\mathbf{f}_i} \in \mathbb{R}^{n_{i} \times n_{{i-1}} }$, $\mathbf{b}_{\mathbf{f}_i} \in \mathbb{R}^{n_{i}}$. Here $\mathbf{v}_{i} \in \mathbb{R}^{n_{i-1}}$ is the input to the $i^{th}$ layer of RNN. See \cite{bMA1c,bMA24} for more information on NN architecture and activation functions. 

The training of RNN is associated with finding the optimal parameters in terms of prediction accuracy. The training data for RNN is defined as in Eq. (\ref{equy}).
For a given initial output and control sequence, the predicted output sequence for the RNN can be computed as:
\begin{equation}
 \label{eqyrnn}
 \begin{aligned}
\hat{\mathbf{Y}}_{\mathrm{D}}&=\left[\begin{matrix} \hat{\mathbf{y}}_{1} & \hat{\mathbf{y}}_{2} & \dots & \hat{\mathbf{Y}}_{\mathrm{D}}\end{matrix}\right]=\left[\begin{matrix} \mathbf{y}_{1}  &  
\mathbf{f}_{\mathrm{NN}}(\mathbf{y}_{1},\mathbf{u}_{1})  & \dots &  
\mathbf{f}_{\mathrm{NN}}(\dots\mathbf{f}_{\mathrm{NN}}(\mathbf{y}_{1},\mathbf{u}_{1}), \mathbf{u}_{{\mathrm{D}-1}})
\end{matrix}\right] =  \mathbf{f}_{\mathrm{yp}}(\bm{\theta}_{\mathbf{f}})
\end{aligned}
\end{equation}
using which the prediction error for RNN is defined as:
\begin{equation}
\label{eqjyrnn}
L_{y_{\mathrm{RNN}}}=\sum_{k=1}^{\mathrm{D}}  {\parallel \mathbf{y}_{k} - \hat{\mathbf{y}}_{k} \parallel }^{2}={\parallel \mathbf{Y}_{\mathrm{D}}-\hat{\mathbf{Y}}_{\mathrm{D}} \parallel }_{F}^{2}=\operatorname{trace}( [\mathbf{Y}_{\mathrm{D}}-\mathbf{f}_{\mathrm{yp}}(\bm{\theta}_{\mathbf{f}})]^{\top}[\mathbf{Y}_{\mathrm{D}}-\mathbf{f}_{\mathrm{yp}}(\bm{\theta}_{\mathbf{f}})])
\end{equation}
which is considered the loss function for the RNN training problem:
\begin{equation}
\label{eqrnnloss}
\underset{\bm{\theta}_{\mathbf{f}}}{\min} \hspace{.2cm}  L_{y_{\mathrm{RNN}}}.
\end{equation} 
The dynamic model identified with RNN can be used for designing output-based NMPC schemes. Given the output reference $\mathbf{y}_{\mathrm{r}},$ the control reference $\mathbf{u}_{\mathrm{r}}$ can be computed by solving the steady-state equation:
\begin{equation} \mathbf{y}_{k+1}=\mathbf{y}_{k} =  \mathbf{y}_{\mathrm{r}}  \implies  \mathbf{y}_{\mathrm{r}}-\mathbf{f}_{\mathrm{NN}}(\mathbf{y}_{\mathrm{r}},\mathbf{u}_{\mathrm{r}};\bm{\theta}_{\mathbf{f}}^{*}) = 0.
\label{eqrnnss}
\end{equation}
Now with $\mathbf{y}_{\mathrm{r}}$ and $\mathbf{u}_{\mathrm{r}},$ the MPC optimization problem can be constructed as in Eq. (\ref{eqnmpc2y}) with $\mathbf{x}_{k|k}$ is replaced by $\mathbf{y}_{k|k}$, and optimized for finding the control sequence.
The algorithm for RNN-based NMPC is summarized below:

\begin{algorithm}[H]
 
	\begin{algorithmic}[1] 
	
	\STATE Require $\mathbf{U}_{\mathrm{D}}$ and $\mathbf{Y}_{\mathrm{D}}$
 \STATE Select  $\mathrm{L},$ $n_{1},\dots,n_{\mathrm{L}}$
 \STATE Initialize $\bm{\theta}_{\mathbf{f}}$
\STATE Compute $\bm{\theta}_{\mathbf{f}}^{*}$ by solving Eq. (\ref{eqrnnloss})   
     \STATE Select $\mathrm{N},$ $\mathbf{y}_{\mathrm{r}},$  $\mathbf{Q}_{\mathbf{y}}$, $\mathbf{R}_{\mathbf{u}}.$ Construct $\mathbf{Q}_{\mathbf{Y}}$ and $\mathbf{R}_{\mathbf{U}}$ as in Eq. (\ref{eqqyru})
     \STATE Compute $\mathbf{u}_{\mathrm{r}}$ by solving Eq. (\ref{eqrnnss})
    \STATE Initialize $\mathbf{U}_{k}$
	\FOR  {$k= 0~to~ \mathrm{N}_{\mathrm{T}}-1 $}
        \STATE Compute $\mathbf{Y}_{r_k}$ using $\mathbf{y}_{\mathrm{r}}$ and $\mathbf{U}_{\mathrm{r}_k}$ using $\mathbf{u}_{\mathrm{r}}$ as in Eq. (\ref{eqqyru}) 
		\STATE  Compute $\mathbf{U}_{k}^{*}$ by solving Eq. (\ref{eqnmpc2y}) with $\mathbf{x}_{k|k}$ is replaced by $\mathbf{y}_{k|k}$
		\STATE Apply  $\mathbf{u}_{k|k}=\mathbf{U}_{k_{(1:m)}}^{*}$ to the system  
		\ENDFOR
    
	\end{algorithmic}
	\caption{: RNN-based D-NMPC}
    \label{algrnn}
\end{algorithm}
\subsubsection{RNN-based D-NMPC: Numerical example}

\begin{figure}[H] 
 		\begin{center}
 		\includegraphics [scale=.1625] {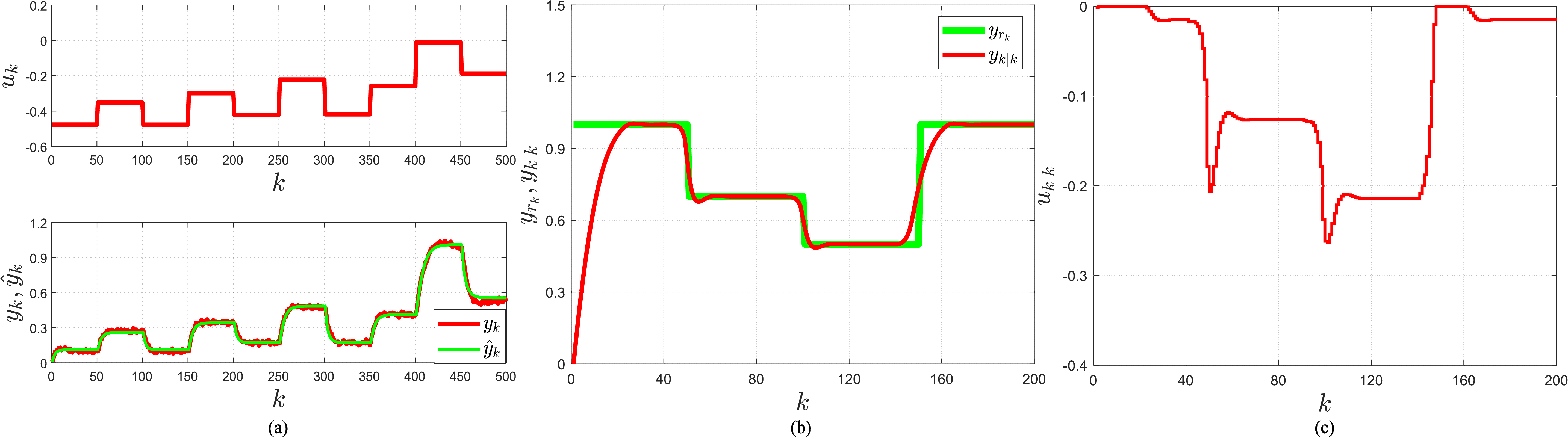}
 		\caption{{ RNN based D-NMPC \hspace{.15cm} (a) Training input and output \hspace{.1cm}(b) Output  with D-NMPC \hspace{.1cm}(c) Control inpput.}}	
        \label{figrnno}
 	\end{center}
 \end{figure}

This section illustrates RNN-based D-NMPC  on a continuous stirred tank reactor (CSTR) system example.    To generate the input-output training data, a second-order state-space model of CSTR is considered \cite{bSJ01}:
\begin{equation}
  \begin{aligned} \mathbf{x}_{k+1}&=\mathbf{x}_{k}+ T {{\left[\begin{matrix} -x_{1_k}+D_{a}(1-x_{1_k})e^{x_{2_k}} \\ x_{2_k}+B\hspace{.1cm}D_{a} (1-x_{1_k})e^{x_{2_k}}-D_{b}(x_{2_k}-u_k) \end{matrix}\right]}}\\ y_{k}&=x_{2_k}+v_{k}
  \end{aligned} 
    \label{eqcstrst}
\end{equation}

where $x_{1_k}$ is the  reactant concentration, $x_{2_k}$ is the  reactant temperature, $u_{k}$ is the  coolant temperature, $y_{k}$ is the measured reactant temperature with $v_{k}$ as the measurement noise,  and $T$ is the sampling period. The dimensionless constants of the system  are chosen as $B=22.0,D_{a}=0.082,D_{b}=3.0.$ 
The input-output data is generated by a forward simulation of Eq. (\ref{eqcstrst}) over $\mathrm{N}_{\mathrm{T}} =1000$ instants with the sampling period $T=1 \hspace{.1cm} sec.,$ initial condition $\mathbf{x}_{0}=\left[\begin{matrix} 0 & 0 \end{matrix}\right]^{\top},$ while the control input $u$ is chosen as a PRMS signal. The first 500 samples of the dataset are used
for training the RNN and the remaining samples for validating the model.
The RNN with $\mathrm{L}=2, n_{1}=5$ and $tanh$ activation function in the hidden layer is trained using the training data, where the loss function is chosen as in Eq. (\ref{eqjyrnn}). The trained RNN model is then used for implementing D-NMPC as in Algorithm \ref{algrnn} with $\mathrm{N}=10$, $\mathbf{Q}_{\mathbf{y}}=2$, and $\mathbf{R}_{\mathbf{u}}=3$.  The response with RNN-based NMPC is shown in Fig. \ref{figrnno}. The training input and output data, along with the output predicted with the trained RNN, are plotted in Fig. \ref{figrnno}(a). The output tracking with RNN-based NMPC is given in Fig. \ref{figrnno}(b), which shows that the output follows the desired reference. Fig. \ref{figrnno}(c) shows the conprol input. 

\subsection{State-space Neural Network based D-NMPC}
State-space neural network (SSNN) is a type of RNN architecture proposed for modeling discrete-time nonlinear systems in state-space form \cite{bJS95,bJZ98,bKK18}. In general, SSNNs are data-driven models trained using the input-output data generated by systems with nonlinear dynamics.  In SSNNs, the model of the system: $\mathbf{f}$ and $\mathbf{h}$ in Eq. (\ref{eqnlfx}) is considered to be unknown. The available information consists of the training input and output as in Eq. (\ref{equy}). 
Then in SSNNs, the objective is to find an approximation to the nonlinear system in Eq. (\ref{eqnlfx}) where the state and output functions are approximated by NNs:
\begin{equation}
  \label{eqnlfxnn}   
 \begin{aligned} \mathbf{x}_{k+1}&=\mathbf{f}_{\mathrm{NN}}(\mathbf{x}_{k},\mathbf{u}_{k};\bm{\theta}_{\mathbf{f}})=
\mathbf{f}_{\mathrm{NN}}(\mathbf{x}_{k},\mathbf{u}_{k}) \\
\hat{\mathbf{y}}_{k}&=\mathbf{h}_{\mathrm{NN}}(\mathbf{x}_{k};\bm{\theta}_{\mathbf{h}})=\mathbf{h}_{\mathrm{NN}}(\mathbf{x}_{k})
 \end{aligned}
\end{equation}
where $\mathbf{x}_{k}\in \mathbb{R}^{l}$ is the predicted state, $\hat{\mathbf{y}}_{k} \in \mathbb{R}^{p}$ is the predicted output,  
$\mathbf{f}_{\mathrm{NN}}:\mathbb{R}^{l} \times \mathbb{R}^{m} \rightarrow \mathbb{R}^{l}$ is the state NN,  $\mathbf{h}_{\mathrm{NN}}:\mathbb{R}^{l} \rightarrow \mathbb{R}^{p}$ is the output NN, and $\bm{\theta}_{\mathbf{f}} \in \mathbb{R}^{n_f},$ and $\bm{\theta}_{\mathbf{h}} \in \mathbb{R}^{n_h}$ contains the parameters (weights and biases) of the state and output NNs, respectively. In the initial versions of SSNN \cite{bJS95,bJZ98,bKK18}, the actual order of the system is assumed to be known, and the order of SSNN is chosen as the system order: $l=n$. However, in general, the order of the system is unknown, and the SSNN order can be different from the actual system order. That's why the notation $\mathbf{x}_{k}$ is used to denote the predicted state with SSNN instead of $\hat{\mathbf{x}}_{k}$. 
Fig. \ref{figSSNN}(b) shows a schematic representation of SSNN. The state and output NNs can be represented as composite functions:
\begin{equation}
 \label{eqfnngnnssnn}
    \begin{aligned}
        & \mathbf{f}_{\mathrm{NN}}=\mathbf{f}_{\mathrm{L}_\mathrm{f}}(\mathbf{f}_{\mathrm{L}_{\mathrm{f}}-1}(\dots(\mathbf{f}_{1}(\mathbf{x}_{k},\mathbf{u}_{k}))))\\
        &\mathbf{h}_{\mathrm{NN}}=\mathbf{h}_{\mathrm{L}_{\mathrm{h}}}(\mathbf{h}_{\mathrm{L}_{\mathrm{h}}-1}(\dots(\mathbf{h}_{1}(\mathbf{x}_{k}))))
    \end{aligned}
\end{equation}
 where $\mathbf{f}_{i}, i=1,2,\dots,\mathrm{L}_\mathrm{f}$ and $\mathbf{h}_{i},i=1,2,\dots,\mathrm{L}_\mathrm{h}$ are layers of the state and output NNs containing $n_{i}$ and $p_{i}$ neurons, respectively,  and can be represented as in Eq. (\ref{eqfigi}).
  Let the training data for SSNN be defined as in Eq. (\ref{equy}).
Given the initial state and control sequence, the predicted state and  output sequence for SSNN become:
\begin{equation}
 \label{eqxnny3}
\begin{aligned}
\mathbf{X}_{\mathrm{D}}&=\left[\begin{matrix} \mathbf{x}_{1} &  \dots & \mathbf{x}_{\mathrm{D}}\end{matrix}\right]=\left[\begin{matrix} \mathbf{f}_{\mathrm{NN}}(\mathbf{x}_{0},\mathbf{u}_{0}) & \dots & \mathbf{f}_{\mathrm{NN}}\big(\dots\mathbf{f}_{\mathrm{NN}}(\mathbf{x}_{0},\mathbf{u}_{0})\dots,\mathbf{u}_{\mathrm{D}-1} \big)\end{matrix}\right]\\
\hat{\mathbf{Y}}_{\mathrm{D}}&=\left[\begin{matrix} \hat{\mathbf{y}}_{1} &  \dots & \hat{\mathbf{y}}_{{\mathrm{D}}}\end{matrix}\right]=\left[\begin{matrix} \mathbf{h}_{\mathrm{NN}}(\mathbf{x}_{1}) &  \dots & \mathbf{h}_{\mathrm{NN}}(\mathbf{x}_{{\mathrm{D}}})\end{matrix}\right] = \mathbf{h}_{\mathrm{NN}}(\mathbf{f}_{\mathrm{NN}}(\mathbf{x}_{0}, \mathbf{U}_{\mathrm{D}})) = \mathbf{f}_{\mathrm{yp}}(\mathbf{x}_{0},\bm{\theta}_{\mathbf{f}},\bm{\theta}_{\mathbf{g}}).
\end{aligned}
\end{equation}
In SSNN, the objective is to find the NN parameters so that the predicted output matches the training output. Consequently, 
the loss function for SSNN is defined as:
\begin{equation}
\label{eqjyssnn}
L_{y_{\mathrm{SSNN}}}=\sum_{k=1}^{\mathrm{D}}  {\parallel \mathbf{y}_{k} - \hat{\mathbf{y}}_{k} \parallel }^{2}={\parallel \mathbf{Y}_{\mathrm{D}}-\hat{\mathbf{Y}}_{\mathrm{D}} \parallel}_{F}^{2}={\parallel \mathbf{Y}_{\mathrm{D}}-\mathbf{f}_{\mathrm{yp}}(\mathbf{x}_{0},\bm{\theta}_{\mathbf{f}},\bm{\theta}_{\mathbf{g}}) \parallel}_{F}^{2}.
\end{equation}
Let ${{ \bm{\theta}=\left[\begin{matrix} \bm{\theta}_{\mathbf{f}} \\ \bm{\theta}_{\mathbf{h}}  \end{matrix}\right]}}$ contains the parameters for the state and output NNs. 
This gives the training problem for SSNN as:
\begin{equation}
\label{eqnlfxnnloss}
\underset{\bm{\theta}}{\min} \hspace{.2cm}  L_{y_{\mathrm{SSNN}}}.
\end{equation} 
The solution for the above optimization problem gives the optimal parameters for $\mathbf{f}_{\mathrm{NN}}$ and $\mathbf{h}_{\mathrm{NN}}$. 
 The SSNN model constructed using the identified parameters from Eq. (\ref{eqnlfsnnloss}) can be used for implementing either output based or state based NMPC schemes. Next, we discuss a D-NMPC scheme that uses a state-space model identified with SSNN for control scheme design. The SSNN model is used for computing the predicted output or state over the prediction horizon. The control sequence is computed to minimize a function of the output/state prediction error. Here we are considering a state-based NMPC for which the state and control reference $\mathbf{x}_{\mathrm{r}},\mathbf{u}_{\mathrm{r}}$ are obtained by solving the steady-state equation:
\begin{equation}
\label{eqref}
\left[\begin{matrix}  \mathbf{y}_{\mathrm{r}}-\mathbf{h}_{\mathrm{NN}}(\mathbf{x}_{\mathrm{r}};\bm{\theta}_{\mathbf{h}}^{*}) \\  \mathbf{x}_{\mathrm{r}}-\mathbf{f}_{\mathrm{NN}}(\mathbf{x}_{\mathrm{r}},\mathbf{u}_{\mathrm{r}};\bm{\theta}_{\mathbf{f}}^{*})    \end{matrix}\right]
=\mathbf{0}.
    \end{equation}
The cost function for state-based NMPC is chosen as in Eq. (\ref{eqmpcjkx}).
 Predicted state and output sequence are defined as in Eq. (\ref{eqxprednl}) where $\mathbf{f},$ $\mathbf{h}$ are replaced with $\mathbf{f}_{\mathrm{NN}},$ $\mathbf{h}_{\mathrm{NN}}$, respectively. 
The first element of $\mathbf{U}_{k}^{*}$ which is $\mathbf{u}_{k|k}^{*}$ is applied to the system at each instant. The algorithm for state-based NMPC using SSNN is given below:

\begin{algorithm}[H]
 
	\begin{algorithmic}[1] 
	
	\STATE Require $\mathbf{U}_{\mathrm{D}}$ and $\mathbf{Y}_{\mathrm{D}}$
 \STATE Select  $\mathrm{L}_{\mathrm{f}},$ $n_{1},\dots,n_{\mathrm{L}_{\mathrm{f}}}$ and $\mathrm{L}_{\mathrm{h}},$ $p_{1},\dots,p_{\mathrm{L}_{\mathrm{h}}}$ 
 \STATE Initialize $\bm{\theta}$
\STATE Compute $\bm{\theta}^{*}$ by solving Eq. (\ref{eqnlfxnnloss})   
     \STATE Select $\mathrm{N},$ $\mathbf{y}_{\mathrm{r}},$  $\mathbf{Q}_{\mathbf{x}}$, $\mathbf{R}_{\mathbf{u}}.$ Construct $\mathbf{Q}_{\mathbf{X}}$ and $\mathbf{R}_{\mathbf{U}}$ as in Eq.  (\ref{eqqyru})
     \STATE Compute $\mathbf{x}_{\mathrm{r}}$ and $\mathbf{u}_{\mathrm{r}}$ by solving Eq. (\ref{eqref}). Compute $\mathbf{X}_{\mathrm{r}}$ using $\mathbf{x}_{\mathrm{r}}$ and $\mathbf{U}_{\mathrm{r}}$ using $\mathbf{u}_{\mathrm{r}}$
    \STATE Initialize $\mathbf{U}_{k}$
    \FOR  {$k= 0~to~ \mathrm{N}_{\mathrm{T}}-1 $}
		\STATE  Compute $\mathbf{U}_{k}^{*}$ by solving Eq. (\ref{eqnmpc2x})
		\STATE Apply  $\mathbf{u}_{k|k}=\mathbf{U}_{k_{(1:m)}}^{*}$ to the system 
		\ENDFOR
	\end{algorithmic}
	\caption{: SSNN-based D-NMPC}
    \label{algssnn}
\end{algorithm}

\subsubsection{SSNN based D-NMPC: Numerical example}
\begin{figure}[H] 
 		\begin{center}
 		\includegraphics [scale=.1625] {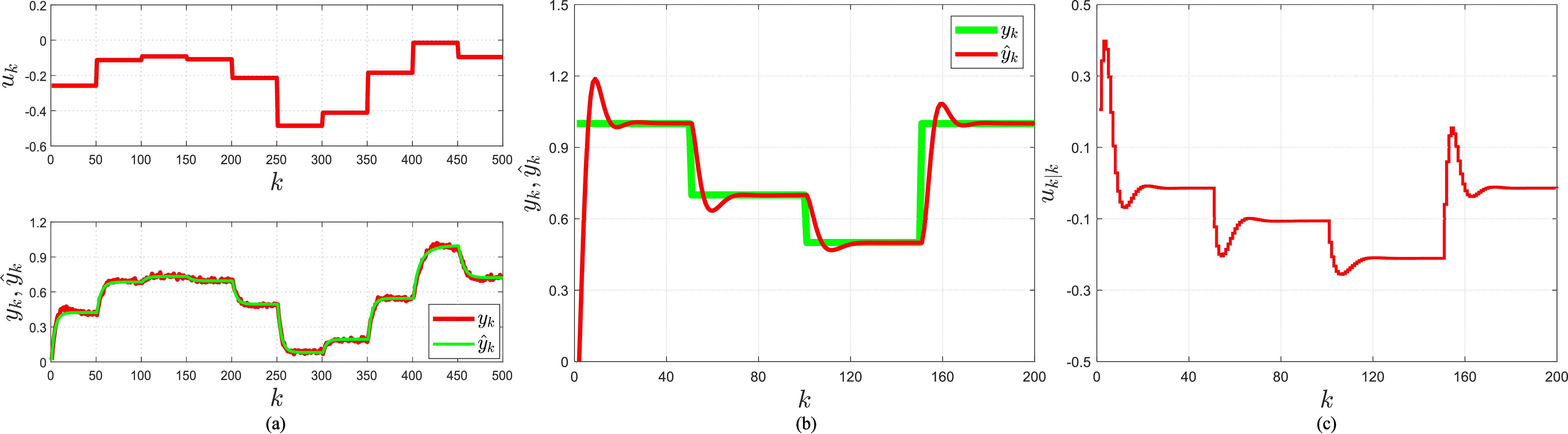}
 		\caption{{ SSNN based D-NMPC \hspace{.15cm} (a) Training input and output \hspace{.1cm}(b) Output  with D-NMPC \hspace{.1cm}(c) Control input.}}	
        \label{figssnny}
 	\end{center}
 \end{figure}

The SSNN-based D-NMPC is illustrated on a CSTR system example.    The CSTR is defined by the discrete-time state equation in Eq. (\ref{eqcstrst}).
The input-output data is generated by a forward simulation of the state equation over $\mathrm{N}_{\mathrm{T}} =1000$ instants with the sampling period $T=1 \hspace{.1cm} sec.,$ and initial condition $\mathbf{x}_{0}=\left[\begin{matrix} 0 & 0 \end{matrix}\right]^{\top}.$ The control input $u$ is chosen as a PRMS signal. The first 500 samples of the dataset are used
for training the SSNN and the remaining samples for
 validation.
 The SSNN with $\mathrm{L}_{\mathrm{f}}=2, n_{1}=5,\mathrm{L}_{\mathrm{h}}=2, p_{1}=5$ and $tanh$ activation function in the hidden layers is selected.
The SSNN is trained using the training data where the loss function is chosen as in Eq. (\ref{eqjyssnn}) and the unconstrained optimization problem in Eq. (\ref{eqnlfxnnloss}) is solved for $\bm{\theta}$ using $fmincon$ in Matlab. The output reference is chosen similarly to the SPC example and the parameters for MPC are chosen as with $\mathrm{N}=10$, $\mathbf{Q}_{\mathbf{x}}={\scriptsize{\left[\begin{matrix} 5 & 0\\ 0 & 4 \end{matrix}\right]}}$, and $\mathbf{R}_{\mathbf{u}}=3$.
The state reference $\mathbf{x}_{\mathrm{r}}$ and control reference $\mathbf{u}_{\mathrm{r}}$ are computed by solving Eq. (\ref{eqref}) with each of the output references. The training input and output data, along with the output predicted with the trained SSNN, are plotted in Fig. \ref{figssnny}(a). The output tracking with SSNN-based NMPC is given in Fig. \ref{figssnny}(b) and the corresponding control input in Fig. \ref{figssnny}(c).   It can be observed that the output converges to the desired reference with the SSNN-based NMPC scheme.
 Note that throughout the paper we denote the output and state predictor function by $\mathbf{f}_{\mathrm{yp}}$ and $\mathbf{f}_{\mathrm{xp}}$, respectively. The predictor function computes the output/state sequence over a prediction horizon $\mathrm{N}$ for a given initial condition and control sequence. Note that the predictor function will be used in both system identification and MPC. In the case of the MPC problem, the predictor function is denoted by  $\mathbf{f}_{\mathrm{yp}}(\mathbf{U}_{k})$ or 
 $\mathbf{f}_{\mathrm{xp}}(\mathbf{U}_{k})$ where $\mathbf{U}_{k}$ is the control input sequence for MPC, which is to be optimized. Similarly, in the case of the SysID problem, the predictor function will be denoted by $\mathbf{f}_{\mathrm{yp}}(\bm{\theta})$ where $\bm{\theta}$ contains all the model parameters to be determined. 
\section{Recent Developements and Future Directions}
In this paper we focused on model-based D-MPC in which a model of the system is first identified from data, then used to predict future system behavior and optimize control actions. While the focus of this paper is on model-based D-MPC, it is worth noting that recent research has also explored model-free data-driven MPC approaches.  One notable approach is DeePC \cite{bJC19,bJB21}, discussed in Section~\ref{secdeepc}, which employs data-driven constraints formulated using input-output Hankel matrices. Several variants of DeePC have also been proposed, including $\gamma$-data driven predictive control, generalized data-driven predictive control, and others \cite{bPV23,bVB23,bML23}.  In these methods, the data itself is directly utilized to compute control schemes, without the need for an explicit system model.

One of the major challenges associated with D-NMPC is that the model order is unknown for many of the practical systems. In the initial approaches of SSNN, the order of the system is assumed to be known, and the SSNN order is chosen as the system order \cite{bJS95,bJZ98,bKK18}. However, in practice, the order of the system is mostly unknown (for systems without any first-principle models available). In \cite{bDM21}, a sparsity-based regularization term is used in an autoencoder for identifying a reduced-order state-space model.  In \cite{bMMS24}, model-order identification problem is addressed where the idea of variance-based ordering \cite{bMPMS24} is used to determine the model order in a systematic way.  The approach is known as \textbf{SSNN with ordered variance} (SSNNO). Variance ordering is achieved in SSNNO by modifying the loss function in Eq. (\ref{eqjyssnn}) as:
\begin{equation}
 \label{eqloss}
L_{y_{\mathrm{SSNNO}}} =  {\underbrace{\alpha_{1}{\parallel \mathbf{Y}_{\mathrm{D}}-\hat{\mathbf{Y}}_{\mathrm{D}} \parallel }_{F}^{2}}_{L_{1}}  +\underbrace{\alpha_{2}  {\parallel \mathbf{W}_{\mathbf{x}}^{\frac{1}{2}}[\mathbf{X}_{\mathrm{D}}-\bar{\mathbf{X}}_{\mathrm{D}}] \parallel }_{F}^{2}}_{L_{2}}+\underbrace{\alpha_{3} {\parallel \bm{\theta}_{\mathbf{f}}  \parallel }_{F}^{2}+\alpha_{4} {\parallel \bm{\theta}_{\mathbf{h}}  \parallel }_{F}^{2}}_{L_{3}}}  
\end{equation}
where $\mathbf{X}_{\mathrm{D}}$ is the predicted state sequence, $\bar{\mathbf{X}}_{\mathrm{D}}$ contains the mean predicted state vector as its elements and $\mathbf{W}_{\mathbf{x}}=diag(w_{1},\dots,w_{l})$ is the weighting matrix used to achieve variance ordering. 
Here, $L_{1}$ is the output prediction error (similar to SSNN) and $L_{2}$ is the state variance  term, which can be rewritten as:
\begin{equation}
    \label{eqj2}
\begin{aligned}   
L_{2}=\alpha_{2} \hspace{.1cm} \operatorname{trace}([\mathbf{X}_{\mathrm{D}}-\bar{\mathbf{X}}_{\mathrm{D}}]^{\top}\mathbf{W}_{\mathbf{x}}[\mathbf{X}_{\mathrm{D}}-\bar{\mathbf{X}}_{\mathrm{D}}])= \alpha_{2}(\mathrm{D}-1)(w_{1}V_{x_{1}}+\dots+w_{l}V_{x_{l}})
   \end{aligned}
    \end{equation}
which is obtained as a weighted sum of the predicted state variances computed over the training data. Consequently, the variance of the state variables can be ordered by adjusting the weighting parameters $w_{i},$ $i=1,2,\dots,l$.  
 Let $s$ denotes the number of state variables in $\mathbf{X}_{\mathrm{D}}$ satisfying:
\begin{equation}
\label{eqvar}
V_{x_{i}}>\delta, \hspace{0.5cm} \delta>0
\end{equation}
which leads to the partition $\mathbf{x} ={{\left[\begin{matrix} \mathbf{x}_{\mathrm{a}} \\ \mathbf{x}_{\mathrm{b}} \end{matrix}\right]}}$, $\mathbf{x}_{\mathrm{a}}\in \mathbb{R}^{s},$ $\mathbf{x}_{\mathrm{b}}\in \mathbb{R}^{l-s}$. The key idea of model order determination/reduction proposed in \cite{bMMS24} is to replace $\mathbf{x}_{\mathrm{b}}$ with the mean vector $\bar{\mathbf{x}}_{\mathrm{b}}$ and reabsorb it in the model as a bias term. This results in $\mathbf{x}_{\mathrm{a}}$ as the only dynamic term in the model. Consequently, the resultant model with $\mathbf{x}_{\mathrm{a}}$ as the state vector will be of order $s$.

\par D-MPC belongs to the broader class of data-driven control and shares common ground with other emerging control techniques, such as adaptive control and \textbf{reinforcement learning} (RL) \cite{bIR11,bRB15}. In particular, D-MPC with online SysID can be viewed as an adaptive control method, where the control input is adjusted based on continuously updated system models estimated in real time. One of the most prominent approaches in model-free control is RL, which has become a rapidly advancing research area in recent years. RL can be viewed as an approximate dynamic programming-based MPC framework, where the cost-to-go function, or value function, is approximated using NNs or other function approximation techniques \cite{bDP19}. In this context, RL uses data-driven methods to iteratively learn optimal control policies without relying on an explicit model of the system dynamics, making it particularly effective for complex, nonlinear, and uncertain systems. The growing interest in RL-based control approaches reflects its potential to solve challenging control problems that conventional model-based methods may struggle with.  Unlike conventional MPC, which requires an accurate model of the system, RL-based MPC adapts its policy based on the feedback obtained from the system's responses. This makes it particularly attractive in scenarios where the model is either unknown, highly complex, or difficult to obtain. The integration of RL with MPC allows for the handling of long-term decision-making while improving the controller's ability to generalize across varying operating conditions.

\section{Conclusions}

This paper discussed various D-MPC approaches, with a particular focus on their numerical implementation. Specifically, it gives a detailed explanation of approaches such as subspace predictive control, PEM-based D-LMPC, RNN, and SSNN-based D-NMPC, etc, with numerical examples. Additionally, the paper highlights current trends in D-MPC research, offering insights into the recent advancements and challenges in the field. The paper also discusses potential future directions for research in the field of D-MPC, considering emerging developments on model-free control algorithms such as reinforcement learning.

\bibliographystyle{unsrt}

\end{document}